\documentclass[journal]{IEEEtran}

\ifCLASSINFOpdf
\else
\fi
\usepackage{amsfonts,amsmath,amssymb}
\usepackage{graphicx}
\usepackage{subfigure}
\usepackage{float}
\usepackage{diagbox}
\usepackage{makecell}
\usepackage{multirow}
\usepackage{booktabs}
\usepackage{amsmath} 
\usepackage[ruled,vlined,algo2e]{algorithm2e}
\usepackage{algorithm}
\usepackage{algorithmic}
\usepackage{verbatim}
\usepackage{xcolor}
\usepackage{array} 

\usepackage{bm}
\usepackage{url} 

\def\dif{\mathop{}\hphantom{\mskip-\thinmuskip}\mathrm{d}}%
\let\daccent\d
\gdef\d{\ifmmode\dif\else\expandafter\daccent\fi}
\begin{document}

\title{Large Language Model Empowered CSI Feedback in Massive MIMO Systems}

\author{Jie Wu,  Wei Xu,~\IEEEmembership{Fellow,~IEEE}, Le Liang,~\IEEEmembership{Member,~IEEE}, Xiaohu You,~\IEEEmembership{Fellow,~IEEE}, and~M{\'e}rouane~Debbah,~\IEEEmembership{Fellow,~IEEE}
\thanks{
Jie Wu, Wei Xu, Le Liang and Xiaohu You are with the National Mobile Communications Research Laboratory, Southeast University, Nanjing 210096, China, and also with the Purple Mountain Laboratories, Nanjing 211111, China (e-mail: 230258127@seu.edu.cn; wxu@seu.edu.cn; lliang@seu.edu.cn; xhyu@seu.edu.cn).

M. Debbah is with the Center for 6G Technology, Khalifa University of Science and Technology, Abu Dhabi, United Arab Emirates (email: merouane.debbah@ku.ac.ae).}
}

\maketitle
\begin{abstract}
Despite the success of large language models (LLMs) across domains, their potential for efficient channel state information (CSI) compression and feedback in frequency division duplex (FDD) massive multiple-input multiple-output (mMIMO) systems remains largely unexplored yet increasingly important. In this paper, we propose a novel LLM-based framework for CSI feedback to exploit the potential of LLMs. We first reformulate the CSI compression feedback task as a masked token prediction task that aligns more closely with the functionality of LLMs. Subsequently, we design an information-theoretic mask selection strategy based on self-information, identifying and selecting CSI elements with the highest self-information at the user equipment (UE) for feedback. This ensures that masked tokens correspond to elements with lower self-information, while visible tokens correspond to elements with higher self-information, thus maximizing the accuracy of LLM predictions. Finally, the LLM leverages its robust modeling capabilities to reconstruct complete CSI representations through contextual inference. This self-information-driven masking strategy integrates the LLM-based masked token prediction mechanism into a coherent, information-driven framework. Numerical results indicate that the proposed LLM-based CSI feedback framework significantly outperforms traditional small models in CSI reconstruction accuracy, leading to substantial improvements in communication rates in multi-user MIMO scenarios. This approach has the potential to address the limitations of CSI reconstruction accuracy that restrict multi-user communication rates. Moreover, the method deploys a lightweight network at the UE, with additional network complexity overhead only at the base station (BS). Finally, the method demonstrates strong generalization across different compression ratios and exhibits excellent transfer learning capabilities across various channel scenarios. These findings pave the way for integrating LLMs into next-generation wireless communication systems.

\end{abstract}
\IEEEoverridecommandlockouts

\IEEEpeerreviewmaketitle
\section{Introduction}

\IEEEPARstart{M}{assive} multiple-input multiple-output (mMIMO) technology has emerged as a foundational enabler of modern wireless communication systems \cite{MIMO_advantage1}, \cite{MIMO_5g}, offering substantial improvements in both spectral efficiency and network capacity by deploying a large number of antennas at the base station (BS) \cite{MIMO_advantage2}. However, the realization of mMIMO performance gains critically depends on the availability of accurate channel state information (CSI) at the transmitter to enable beamforming design. While in time division duplex (TDD) systems CSI can be directly acquired by leveraging channel reciprocity, it must be estimated at the user equipment (UE) and then fed back to the BS \cite{feedback1} in frequency division duplex (FDD) systems. As the number of antennas scales up in mMIMO deployments, the overhead of downlink CSI feedback in mMIMO FDD systems becomes prohibitively large, imposing significant feedback challenges at UE \cite{feedback2}. 

To mitigate the substantial CSI feedback overhead inherent in FDD mMIMO systems, a wide range of techniques have been developed, with particular emphasis on leveraging both spatial and temporal correlations in wireless propagation channels. Early works utilized compressed sensing \cite{MIMO_CS}, sinusoidal basis modeling \cite{MIMO_compression1}, and eigenvalue decomposition \cite{MIMO_compression2} for efficient CSI compression representation. Nonetheless, these approaches typically result in a feedback overhead that increases linearly with the number of transmit antennas, thereby limiting their scalability in practical large-scale deployments. Alternatively, vector and matrix codebook-based approaches have been adopted, including the Type I/II codebooks in 5G NR \cite{codebook1}, angle-domain sparse codebook constructions \cite{codebook2}, and adaptive subspace quantization schemes \cite{codebook3}. However, codebook-based methods require careful design tailored to specific channel conditions, resulting in limited flexibility and poor adaptability to diverse wireless environments.

Benefiting from the powerful parallel computing and feature extraction capabilities of deep learning (DL), DL-based CSI compression feedback methods have been extensively studied and have demonstrated significantly superior performance compared to traditional approaches. One of the early works \cite{deepcsi1} treated the sparse angle-delay domain CSI as an image and applied an autoencoder to perform compression and reconstruction. Subsequent studies have focused on improving performance by designing more advanced neural networks (NNs). For instance, CRNet \cite{deepcsi2} and ACRNet \cite{deepcsi3} employed convolutional neural networks (CNNs) with varying kernel sizes to simultaneously capture the sparse and dense features of the CSI. CLNet \cite{deepcsi4} introduced an attention mechanism to enhance feature representation, while study \cite{deepcsi5} proposed a hybrid architecture that combines CNNs with weighted attention to separately extract Line-of-Sight (LoS) and non-Line-of-Sight (nLoS) components. To further reduce information redundancy in CSI compression, study \cite{deepcsi6} introduced a self-information based CSI compression model inspired by image information theory. In addition, temporal correlations in time-varying CSI were exploited by leveraging recurrent neural networks (RNNs) \cite{deepcsi7} and long short-term memory (LSTM) modules \cite{deepcsi8}, leading to improved reconstruction accuracy across successive channel states. In study \cite{TransInDecNet}, the authors designed a novel CSI feedback network called TransInDecNet based on the transformer architecture. Furthermore, significant research has been dedicated to developing specialized lightweight architectures to minimize computational complexity at the UE side. For instance, works such as \cite{lightcsi1} and \cite{lightcsi2} focused on optimizing network structure through techniques like weight sharing and component replacement to reduce model size and inference time, acknowledging the critical challenge of deploying efficient deep learning models in resource-constrained networks \cite{jcyao}. Further in \cite{deepcsi9}, the implicit reciprocity between downlink and uplink CSI in FDD systems was successfully exploited to reduce feedback overhead. Beyond the digital compression paradigm, another active research direction is superimposed CSI feedback, which is a spectrum-efficient scheme that overlays CSI signals onto uplink data transmission. Due to its high spectral efficiency, superimposed CSI feedback is particularly well-suited for resource-constrained scenarios such as massive machine-type communication or dense Internet of Things (IoT) deployments. Recent works have explored different superimposed CSI feedback strategies to minimize interference and maximize spectral efficiency, as detailed in \cite{scsi1-1}, \cite{scsi1-2}, and \cite{scsi1-3}.

Although existing DL-based CSI feedback methods have achieved promising reconstruction performance under relatively simple channel conditions and high compression ratios, their performance degrades significantly as the channel environment becomes more complex or the compression ratio decreases. On one hand, conventional small DL models exhibit limited decoding and prediction capabilities, making it difficult to recover the complex high-dimensional mMIMO CSI matrix from a small number of feedback codewords \cite{llmstrong}. This limitation necessitates the introduction of large-capacity architectures to model the complex, non-linear channel distributions beyond the reach of conventional small-scale networks. On the other hand, the image structure within the latent space introduces additional structural redundancy, which prevents feedback codewords from fully capturing the underlying CSI information \cite{1dtoken}. The resulting reconstruction error creates a precision bottleneck that urgently needs to be overcome to unlock the full throughput potential of multi-user massive MIMO systems. These challenges have become a key bottleneck for improving the performance of deep mMIMO CSI feedback. 

The emergence of large language models (LLMs) has revolutionized traditional deep learning approaches in the field of artificial intelligence (AI), demonstrating superior capabilities in sequence prediction, feature extraction, and generalization compared to conventional small models \cite{llmstrong}, \cite{ILAC}. Beyond prediction, the extensive pre-training of LLMs provides them with remarkable robustness, adaptability, and an inherent capacity to capture complex, non-linear patterns, making them highly suitable for challenging communication tasks \cite{llmsum1}, \cite{llmsum2}. For example, specific prompts can be designed to guide LLMs in generating domain-specific knowledge for wireless communication \cite{llmopt1}, \cite{llmopt2}, \cite{llmopt3}. LLMs are also being utilized for communication-related prediction tasks, leveraging their sequential prediction capabilities \cite{timellm}, \cite{llmpred1}, and have been shown to produce high-accuracy CSI predictions \cite{llmpred2}, \cite{llmpred3}. Furthermore, to leverage the capability of LLMs in handling multiple tasks concurrently, the authors in \cite{llmmult1} and \cite{llmmult2} proposed a multi-task LLM framework that enables simultaneous execution of multiple communication tasks. The demonstrated success of LLMs across numerous wireless communication tasks motivates us to explore their potential for deep CSI feedback. This motivation is further reinforced by the theoretical equivalence between compression and prediction tasks \cite{preiscompress}. However, existing LLM-based approaches for CSI compression and feedback remain in their infancy \cite{llmcsi1}. 

A fundamental challenge in applying LLMs to CSI compression lies in their architectural nature: LLMs are not autoencoders and inherently lack the capability for compression and decompression, as their inputs and outputs typically maintain equal lengths \cite{attention}. Moreover, the absence of pretraining for compression-related tasks makes it difficult for LLMs to reconstruct the original CSI from compressed codewords, necessitating the design of external encoders and decoders. In contrast, masked token prediction is a well-established pretrained task in most LLMs. In this task, a portion of the input token sequence is masked, and the model is trained to predict the masked tokens based on the visible tokens. Due to extensive pretraining, LLMs are highly proficient at this task. Furthermore, LLMs, as transformer-based models, possess strong contextual modeling capabilities, enabling them to capture long-range dependencies and seamlessly integrate contextual cues from visible tokens to infer masked ones. 

From an information theory perspective \cite{informationtheory}, masked token prediction can be viewed as analogous to a compression task: visible tokens represent the compressed codewords, while masked tokens represent the discarded information. This approach leverages the universal sequence modeling capability of the transformer architecture, which excels at identifying long-range dependencies common to both linguistic and physical data. Based on this insight, we propose a novel framework that reformulates CSI compression as a masked token prediction problem. CSI elements with high information are selected and fed back to BS, while low-information CSI elements are masked. At the BS, the CSI matrix is then divided into small patches, each treated as a CSI token. CSI tokens containing high-information CSI elements received from UE are considered visible tokens, whereas those comprising low-information elements not fed back are treated as masked tokens. The LLM then predicts masked tokens based on the visible tokens. Inspired by our previous work \cite{deepcsi6}, we quantify the information content of each CSI element by measuring its variation relative to neighboring elements, a metric referred to as self-information. CSI elements that exhibit larger relative changes are considered carrying higher self-information, therefore being prioritized for feedback at UE and leading to improved CSI reconstruction accuracy.

In this paper, we propose a novel LLM-based deep CSI feedback framework, referred to as LLMCsiNet. The proposed LLMCsiNet reformulates the CSI compression and feedback task as a masked token prediction task to enable accurate CSI reconstruction with a limited number of feedback codewords. The main contributions of this work are summarized as follows.
\begin{figure*}[t]
	\centering 
	\includegraphics[width=\linewidth]{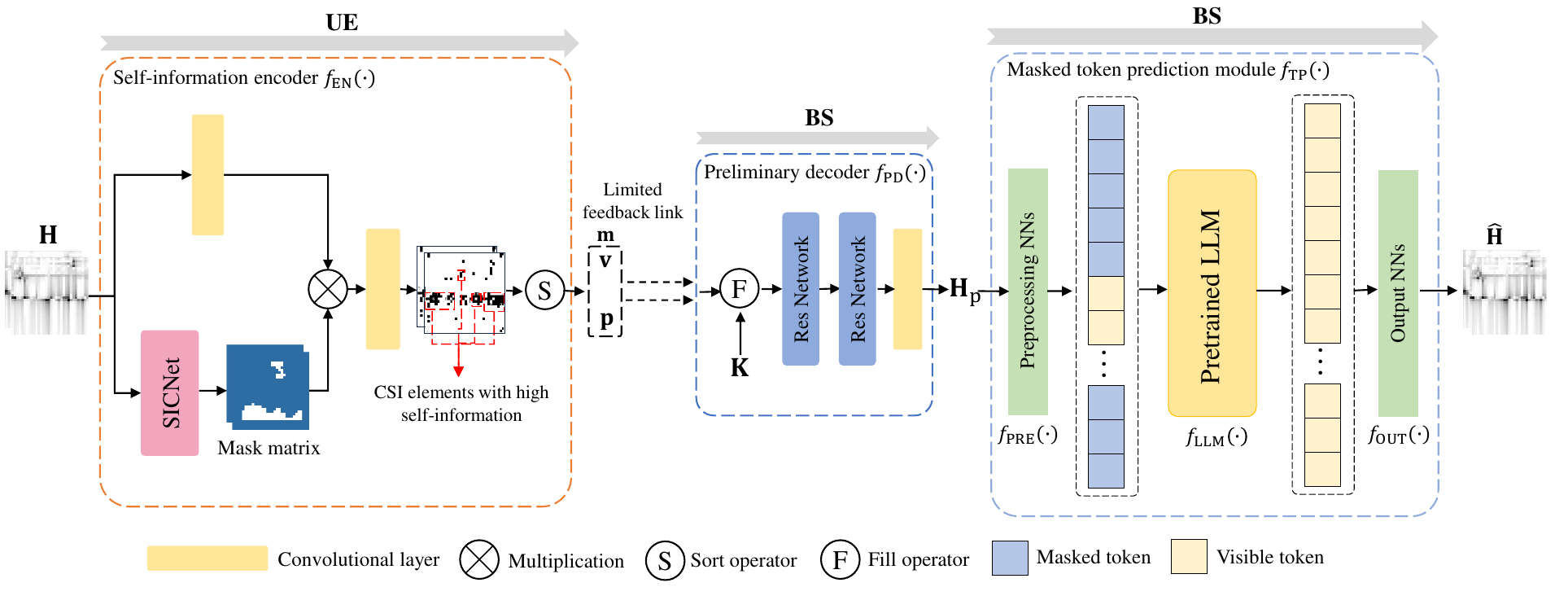}
	\caption{The proposed LLMCsiNet framework for CSI compression and feedback.}
	\label{framework}
\end{figure*}

\begin{itemize}
\item[$\bullet$]We introduce a novel framework for CSI feedback by reformulating the task as a masked CSI token prediction problem. This fundamentally shifts the paradigm from traditional compression to context-aware reconstruction. Specifically, the framework utilizes a self-information based mechanism to select and feed back the most critical CSI elements as visible tokens. The core mechanism then harnesses the predictive and contextual modeling capabilities of the LLM at the BS to intelligently infer and reconstruct the vast majority of the missing masked tokens. This approach significantly mitigates the severe performance degradation commonly observed in conventional small models under complex channel conditions and ultra-low compression ratios.
\item[$\bullet$]To flexibly control the model complexity, only a subset of transformer layers from the LLM is used rather than the full LLM network. Most of the network complexity introduced by the LLM is handled at the BS, where hardware resources are more abundant, while the UE only needs to run a lightweight module with lower network complexity than many small model-based methods, e.g., \cite{deepcsi1}, \cite{deepcsi2}.
\item[$\bullet$]Numerical results show that the proposed LLMCsiNet yields a significant performance gain of $3$ to $10$ dB over state-of-the-art small models under different channel conditions and compression settings depending on the model scale of LLMs. In addition, LLMCsiNet is capable of handling feedback codewords at multiple compression ratios simultaneously, outperforming small models that are specifically trained for each individual compression ratio. Furthermore, LLMCsiNet achieves the same accuracy as small models trained on full datasets by leveraging transfer learning with a small number of samples. 
\item[$\bullet$]This method allows for the reconstruction of CSI at the BS with extremely high precision, enabling communication rates in multi-user MIMO scenarios that far exceed traditional small model methods. It has the potential to overcome the current limitations and bottlenecks in multi-user MIMO transmission performance, which are largely constrained by the accuracy of CSI reconstruction at BS. 
\end{itemize}

The remainder of this paper is organized as follows. Section~\uppercase\expandafter{\romannumeral2} introduces the system model of deep CSI feedback. Section~\uppercase\expandafter{\romannumeral3} introduces the design motivation and self-information, while Section~\uppercase\expandafter{\romannumeral4} presents the detailed design and training procedure of the proposed LLMCsiNet. Numerical results are provided in Section~\uppercase\expandafter{\romannumeral5}, and conclusions are drawn in Section~\uppercase\expandafter{\romannumeral6}.

\emph{Notations}: Throughout this paper, scalar variables are denoted by normal-face letters, whereas column vectors and matrices are respectively represented by boldface lowercase and uppercase letters. The expectation operator is expressed as $\mathbb{E} \left[ \cdot \right]$, and the matrix space of dimension $M \times N$ is denoted by $\mathbb{C}^{M \times N}$. The Hermitian transpose is denoted by the superscript $(\cdot)^H$. The Euclidean norm is represented by $\left\| \cdot \right\|_2$. Additionally, the trace of a matrix is denoted by $\mathrm{Tr}(\cdot)$

\section{System Model}
In this paper, we consider an FDD mMIMO system in which the BS and UE are equipped with $N_\mathrm{t}$ antennas and one antenna, respectively. Orthogonal frequency-division multiplexing (OFDM) with $N_\mathrm{f}$ subcarriers is employed. The received signal $y_n \in \mathbb{C}$ at a UE on the $n$-th subcarrier is represented as:
\begin{equation}
{y_n}={\mathbf{h}_n^H} \mathbf{w}_n {x_n}+{z}_n, \quad n=1, \ldots, N_\mathrm{f},
\end{equation}
where $\mathbf{h}_n \in \mathbb{C}^{N_\mathrm{t} \times 1}$ denotes the channel vector at the $n$-th subcarrier, $\mathbf{w}_n \in \mathbb{C}^{N_\mathrm{t} \times 1}$ represents the corresponding beamformer, and $x_n \in \mathbb{C}$ and $z_n \in \mathbb{C}$ are the transmitted signal and the additive noise, respectively. By stacking the channel vectors across all subcarriers, we obtain the channel matrix ${\widetilde{\mathbf{H}}} \in \mathbb{C}^{N_\mathrm{f} \times N_\mathrm{t}}$, defined as ${\widetilde{\mathbf{H}}}=\left[\mathbf{h}_1, \mathbf{h}_2, \ldots,\mathbf{h}_{N_\mathrm{f}}\right]^H$. To reduce the CSI feedback overhead, a two-dimensional (2D) discrete Fourier transform (DFT) is applied to convert the CSI from the spatial domain to the angle-delay domain \cite{Hdelay}, resulting in
\begin{equation}
{\overline{\mathbf{H}}}=\mathbf{F}_\mathrm{c} \widetilde{\mathbf{H}} \mathbf{F}_\mathrm{d},
\end{equation}%
\noindent where $\mathbf{F}_\mathrm{c} \in \mathbb{C}^{N_\mathrm{f} \times N_\mathrm{f}}$ and $\mathbf{F}_\mathrm{d} \in \mathbb{C}^{N_\mathrm{t} \times N_\mathrm{t}}$ denote DFT matrices of appropriate dimensions. ${\overline{\mathbf{H}}} \in \mathbb{C}^{N_\mathrm{f} \times N_\mathrm{t}}$ denotes the CSI matrix in the angle-delay domain. In practical wireless environments, physical channels exhibit finite multipath delay intervals, which implies that ${\overline{\mathbf{H}}}$ contains nonzero elements only within limited delay intervals. To take advantage of this structured sparsity, we truncate ${\overline{\mathbf{H}}}$ by retaining only the first $N_\mathrm{c}$ rows that contain significant values, yielding a truncated matrix denoted by $\mathbf{H}_\mathrm{c} \in \mathbb{C}^{N_\mathrm{c} \times N_\mathrm{t}}$. Since typically $N_\mathrm{c} \ll N_\mathrm{f}$, this truncation significantly reduces the number of feedback parameters from $N_\mathrm{f} \times N_\mathrm{t}$ to $N_\mathrm{c} \times N_\mathrm{t}$ \cite{Hdelay}. Furthermore, as NNs operate on real-valued data while CSI is complex-valued, we transform $\mathbf{H}_\mathrm{c}$ into a real-valued matrix $\mathbf{H} \in \mathbb{C}^{2 \times N_\mathrm{c} \times N_\mathrm{t}}$ suitable for NNs’ input.

To further compress $\mathbf{H}$, we design an LLM-based CSI feedback framework, called LLMCsiNet, as depicted in Fig.~ \ref{framework}. An encoder based on self-information maximization is deployed at UE to compress $\mathbf{H}$ into a codeword of specific length. The encoding process can be expressed as
\begin{equation}
\mathbf{m}=f_{\mathrm{EN}}(\mathbf{H}),
\end{equation}
where $\mathbf{m} \in \mathbb{C}^{M \times 1}$ denotes the codeword, commonly $M \ll N_\mathrm{c} \times N_\mathrm{t}$, and $f_\mathrm{EN}(\cdot)$ represents the encoder. The corresponding decoding process at BS is expressed as
\begin{equation}
\mathbf{\widehat{H}}=f_{\mathrm{TP}}(f_{\mathrm{PD}}(\mathbf{m})),
\end{equation}
where $\mathbf{\widehat{H}} \in \mathbb{C}^{2 \times N_\mathrm{c} \times N_\mathrm{t}}$ denotes the reconstructed CSI matrix, and $f_\mathrm{PD}(\cdot)$ represents a preliminary decoding network responsible for the initial reconstruction of CSI. The $f_\mathrm{TP}(\cdot)$ module refers to the LLM-based masked token prediction module, which performs final reconstruction by predicting the masked CSI tokens.

\section{Design Motivation and Self-information}
A major challenge in applying LLMs to the CSI compression and feedback task lies in the fact that LLMs are not inherently designed for compression or decompression because their input and output dimensions are typically identical. Specifically, an LLM accepts a sequence of $L$ tokens as input, where each token corresponds to a word in natural language. The LLM processes this input and produces an output sequence of the same length. During the pretraining phase of LLMs, a commonly adopted strategy is to mask a subset of tokens within a sentence, referred to as masked tokens, while keeping the remaining tokens unchanged, known as visible tokens. The LLMs are then trained to predict the masked tokens based on the visible tokens. From an information-theoretic perspective, this context-based masked token prediction can be regarded as an asymmetric form of information compression. 

Conventional small models for CSI compression typically aim to achieve globally optimal reconstruction by encoding the entire CSI into a low-dimensional codeword through a data-driven compression network. However, this global compression strategy is fundamentally misaligned with the operational mechanism of LLMs. To address this mismatch, we propose a novel framework that reformulates CSI compression and reconstruction as a masked token prediction task. Although pre-trained on text, the LLM is repurposed here because the sequential structural correlations in CSI are mathematically isomorphic to the grammatical dependencies in language. During the encoding stage at UE, only a subset of locally significant CSI elements is fed back. The received CSI elements at BS are treated as visible tokens, while the remaining untransmitted elements are regarded as masked tokens, which are then inferred by the LLM based on the visible tokens. In this framework, the encoder is only responsible for extracting and feeding back locally significant CSI elements, rather than encoding the full CSI. This framework leads to a fundamental design consideration regarding the selection of CSI elements to be fed back to the LLM. From the perspective of LLMs, the higher the information content carried by the visible tokens in the input sequence, the more accurately the model can predict the masked tokens. This observation highlights the necessity of a principled method to quantify the information content of CSI elements. It is desirable to prioritize the feedback of high-information components while omitting low-information ones, which can be reliably inferred by the LLM.

Inspired by our previous work \cite{deepcsi6}, we introduce self-information as a metric to quantify the information of individual CSI elements. This metric serves as a guide for the encoder in selecting the most informative CSI elements for feedback. For a given channel matrix $\mathbf{H} \in \mathbb{C}^{2 \times N_\mathrm{c} \times N_\mathrm{t}}$, we define the $j$-th CSI element as $c_j$. If a CSI element exhibits significant variation relative to its neighboring elements, it is usually unpredictable and considered to possess high self-information. In contrast, if the element shows minimal variation compared to its neighbors and can be easily inferred, it is regarded as an element of low self-information. According to information theory \cite{informationtheory}, the self-information of each CSI element is expressed as
\begin{equation}
I_j = -\log_2 q_j, \quad \forall j,
\label{selinfo}
\end{equation}
where $q_j$ denotes the probability of $c_j$ and $\log_2$ denotes the base-2 logarithm. Intuitively, a CSI element with low probability $q_j$ corresponds to a highly informative element of CSI, while elements with high probability can be considered as redundant and contribute marginally to CSI reconstruction.
\begin{figure*}[t]
	\centering 
	\includegraphics[width=\linewidth]{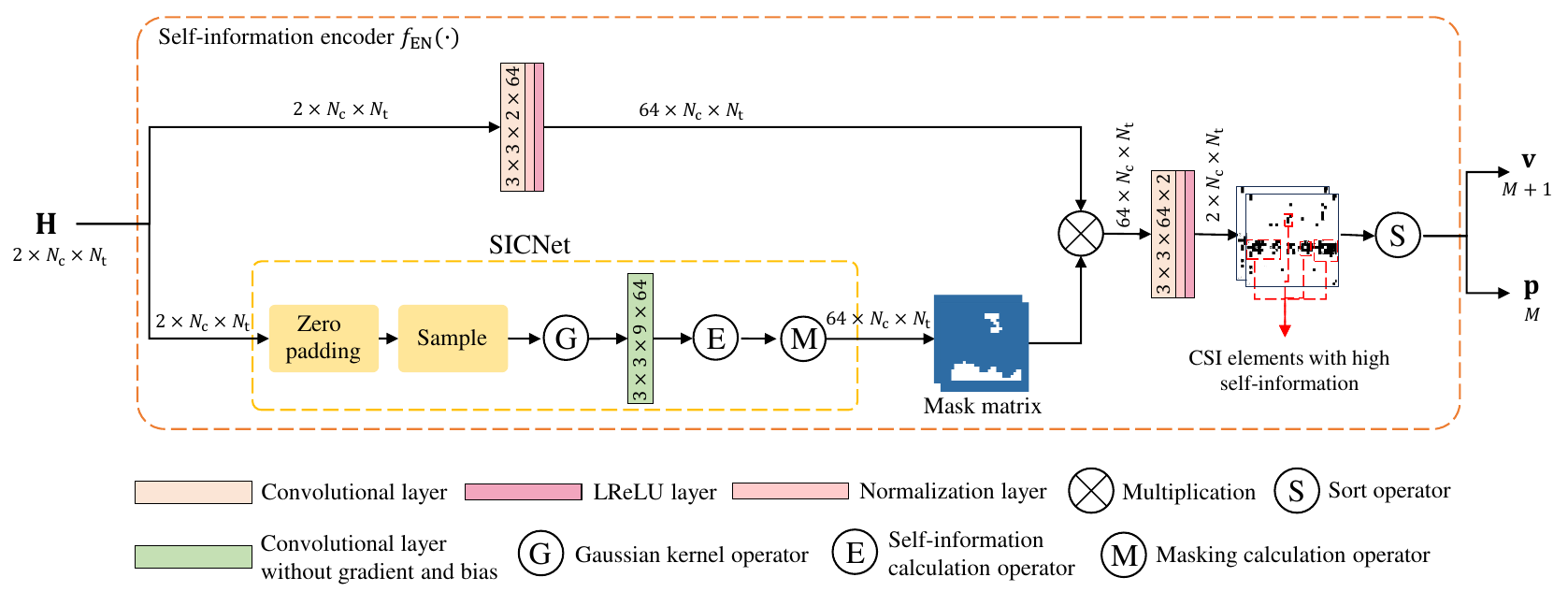}
	\caption{Design details of the self-information encoder $f_{\mathrm{EN}}(\cdot)$.}
	\label{siencoder}
\end{figure*}

To estimate $q_j$, we consider a local neighborhood $\mathcal{N}_j$ centered at $c_j$, determined by a Manhattan radius $R$, which contains a total of $(2R+1)^2$ elements. The assumption that all neighboring elements $c_{j,r}'$ follow the same distribution as $c_j$ is justified by the strong spatial and spectral correlation of the CSI features, which ensures local statistical stationarity. we apply a kernel density estimation (KDE) method to obtain the estimated probability $\hat{q}_j$
\begin{equation}
\hat{q}_j = \frac{1}{(2R+1)^2} \sum_{c_{j,r}' \in \mathcal{N}_j} K(c_j, c_{j,r}'),
\end{equation}
where $K(\cdot, \cdot)$ is a Gaussian kernel with bandwidth $h$
\begin{equation}
K(c_j, c_{j,r}') = \frac{1}{\sqrt{2\pi} h} \exp\left(-\frac{| c_j - c_{j,r}' |^2}{2h^2}\right).
\label{gauker}
\end{equation}
Substituting into (\ref{selinfo}) yields
\begin{equation}
\hat{I}_j = -\log_2 \left( \frac{1}{(2R+1)^2} \sum_{c_{j,r}'\in \mathcal{N}_j} \frac{1}{\sqrt{2\pi} h} e^{\left(-\frac{| c_j - c_{j,r}' |^2}{2h^2} \right)} \right) + \mathrm{C},
\end{equation}
where $\mathrm{C}$ is a constant term. Computing the Gaussian kernel between a CSI element and all elements within its $\mathcal{N}_j$ can incur significant computational overhead, especially when the Manhattan radius $R$ is large.
Therefore, in practical implementation, we uniformly select 9 elements from $\mathcal{N}_j$ to estimate the self-information of $c_j$. It yields               
\begin{equation}
\hat{I}_j = -\log_2 \left( \frac{1}{9} \sum \frac{1}{\sqrt{2\pi} h} e^{\left(-\frac{| c_j - c_{j,r}' |^2}{2h^2} \right)} \right) + \mathrm{C}.
\label{info2}
\end{equation}
After computing the self-information of all CSI elements, we selectively discard elements with low information, enabling the model to focus on critical elements and thereby achieve more efficient compression.

\section{Proposed Architecture of LLMCsiNet}
In this section, we propose an LLM-based CSI feedback framework that fully leverages the potential of LLMs by reformulating the CSI compression and reconstruction task as a masked token prediction problem, while simultaneously maximizing the self-information contained in the compressed codewords. The goal of this framework is to harness the powerful contextual modeling and predictive abilities of LLMs to achieve decoding performance that surpasses that of smaller models, enabling substantially more accurate CSI reconstruction with limited feedback overhead. The design and implementation details of the proposed framework are elaborated in the following.

\subsection{Network Framework}
Although existing small-model CSI compression and feedback methods based on auto-encoder architectures have demonstrated superior performance compared to traditional approaches, their reconstruction accuracy degrades significantly in complex channel environments especially when the compression ratio decreases. This limitation arises from the inherent capacity constraints of small models. To address this issue, we propose LLMCsiNet, a novel CSI compression and feedback framework based on LLMs as presented in Fig.~\ref{framework}. The framework of LLMCsiNet comprises three network modules: $f_{\mathrm{EN}}(\cdot)$, $f_{\mathrm{PD}}(\cdot)$, and $f_{\mathrm{TP}}(\cdot)$. In the following, we introduce the three modules of LLMCsiNet.

\begin{figure*}[t]
	\centering 
	\includegraphics[width=\linewidth]{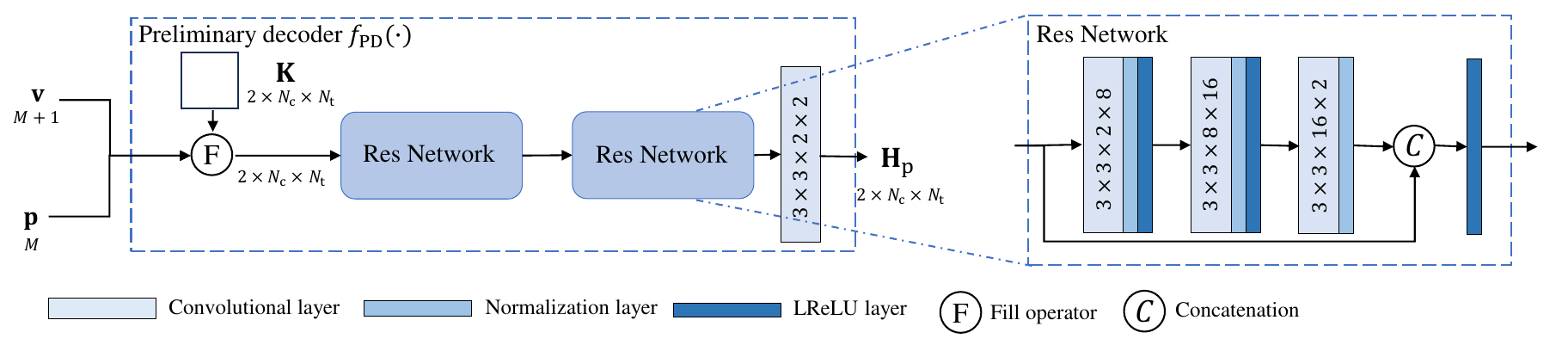}
	\caption{Design details of the preliminary decoder $f_{\mathrm{PD}}(\cdot)$.}
	\label{fdecoder}
\end{figure*}
\textit{Self-information encoder} $f_{\mathrm{EN}}(\cdot)$: This module is designed to be lightweight and is deployed at the UE to compress the CSI matrix $\mathbf{H}$. As depicted in Fig.~\ref{framework}, this module takes $\mathbf{H}$ as input and first extracts a CSI feature map using a convolutional layer. Then, a self-information computation network (SICNet) estimates the self-information of CSI elements by (\ref{info2}) and produces a mask matrix. The mask matrix is subsequently multiplied with the CSI feature map to suppress elements with low self-information. The masked feature map is subsequently fed into another convolutional layer to generate codewords. The codewords are then sorted by a sort operator, and the top $M$ elements with the highest self-information are selected. The corresponding value vector $\mathbf{v} \in \mathbb{C}^{M+1}$, which includes an additional element to transmit the mean value, and position index vector $\mathbf{p} \in \mathbb{C}^{M}$ are fed back to the BS. 

\textit{Preliminary decoder} $f_{\mathrm{PD}}(\cdot)$: This module is deployed at the BS and performs a preliminary reconstruction of the CSI. As presented in Fig.~\ref{framework}, the module first generates a matrix $\mathbf{K} \in \mathbb{C}^{2 \times N_\mathrm{c} \times N_\mathrm{t}}$, where all elements are initialized to the mean value of the original CSI matrix $\mathbf{H}$. The top 
$M$ CSI elements with the highest self-information in $\mathbf{v}$ are then inserted into their corresponding positions in 
$\mathbf{K}$ according to the position indices in $\mathbf{p}$. The inserted matrix is subsequently refined through two cascaded residual networks (Res Networks) to output the preliminary reconstructed CSI $\mathbf{H}_\mathrm{p}$.

\textit{Masked token prediction module} $f_{\mathrm{TP}}(\cdot)$: This module, deployed at the BS, serves as the core component of the proposed LLMCsiNet. As depicted in Fig.~\ref{framework}, it reformulates the CSI reconstruction task as a masked token prediction problem and consists of three submodules: the preprocessing NNs $f_\mathrm{PRE}(\cdot)$, the pretrained LLM $f_\mathrm{LLM}(\cdot)$, and the output NNs $f_\mathrm{OUT}(\cdot)$. The preliminarily reconstructed CSI from $f_{\mathrm{PD}}(\cdot)$ is first fed into $f_\mathrm{PRE}(\cdot)$ for preprocessing. Notably, $f_\mathrm{LLM}(\cdot)$ consists only of the transformer layers from a pretrained LLM, excluding the original embedding and output layers. Finally, the predicted CSI token sequence is passed through $f_\mathrm{OUT}(\cdot)$, thereby producing the final reconstructed CSI.

The following subsections present a detailed exposition of the design principles and architectures of $f_{\mathrm{EN}}(\cdot)$, $f_{\mathrm{PD}}(\cdot)$, and $f_{\mathrm{TP}}(\cdot)$, highlighting the underlying concepts and specific implementation details of each module.

\subsection{Self-information Encoder $f_{\mathrm{EN}}(\cdot)$}
Previous DL-based CSI feedback methods typically aim to maximize the reconstruction accuracy of the entire CSI matrix during encoding. In contrast, the proposed self-information encoder $f_{\mathrm{EN}}(\cdot)$ focuses on maximizing the reconstruction accuracy in the most informative regions, while intentionally ignoring regions with low self-information. 

The detailed architecture of $f_{\mathrm{EN}}(\cdot)$ is depicted in Fig.~\ref{siencoder}. It primarily consists of two convolutional layers and a SICNet. In the convolutional layers, the four numbers represent the kernel height, the kernel width, the number of input feature maps, and the number of output feature maps, respectively. Both convolutional layers employ batch normalization (BN) to enhance training stability and adopt a LeakyReLU (LReLU) activation function to introduce nonlinearity into the network. The LReLU function is defined as
\begin{equation}
\text{LReLU}(x) = 
\begin{cases}
x, & \text{if } x \geq 0 \\
0.3x, & \text{otherwise}.
\label{lrelu}
\end{cases}    
\end{equation}

The input CSI $\mathbf{H}$ is simultaneously fed into a convolutional layer and the SICNet. The convolutional layer extracts CSI feature maps, while the SICNet generates a binary mask matrix composed of $0$ and $1$. In this mask matrix, positions corresponding to high self-information elements of $\mathbf{H}$ are set to $1$, while other positions with low self-information are set to $0$. The mask matrix is then element-wise multiplied with the CSI feature map to perform informative dropout. Upon receiving $\mathbf{H}$, the SICNet first applies zero padding to expand the dimensions of $\mathbf{H}$ to $2 \times (N_\mathrm{c}+R-1) \times (N_\mathrm{t}+R-1)$, where $R$ denotes the Manhattan radius of $\mathcal{N}_j$. The zero-padding ensures that every element in $\mathbf{H}$ has access to a neighborhood of consistent size. Subsequently, $9$ elements are uniformly sampled from each $\mathcal{N}_j$, and a Gaussian kernel operator is applied to compute the distances between the center element and each sampled element according to equation (\ref{gauker}), resulting in $9$ Gaussian kernel feature maps. These maps are then processed by a convolutional layer without gradients and biases to produce $64$ feature maps. The self-information calculation operator then calculates $64$ self-information maps based on equation (\ref{info2}). Following this, a masking calculation operator produces a binary mask matrix of size $64 \times N_\mathrm{c} \times N_\mathrm{t}$, where elements with self-information below a threshold $T$ are set to $0$, and the remaining elements are set to $1$, effectively eliminating redundant information from the CSI feature maps. 

Finally, the filtered CSI feature map is passed through another convolutional layer to produce a self-information image with the same dimensionality as $\mathbf{H}$. A sort operator is then applied to extract the $M$ most informative values along with their corresponding position indices from the self-information image. These are output as vectors $\mathbf{v}$ and $\mathbf{p}$, which will be used for CSI reconstruction at the decoder. 

\subsection{Preliminary Decoder $f_{\mathrm{PD}}(\cdot)$}
\begin{figure*}[t]
	\centering 
	\includegraphics[width=\linewidth]{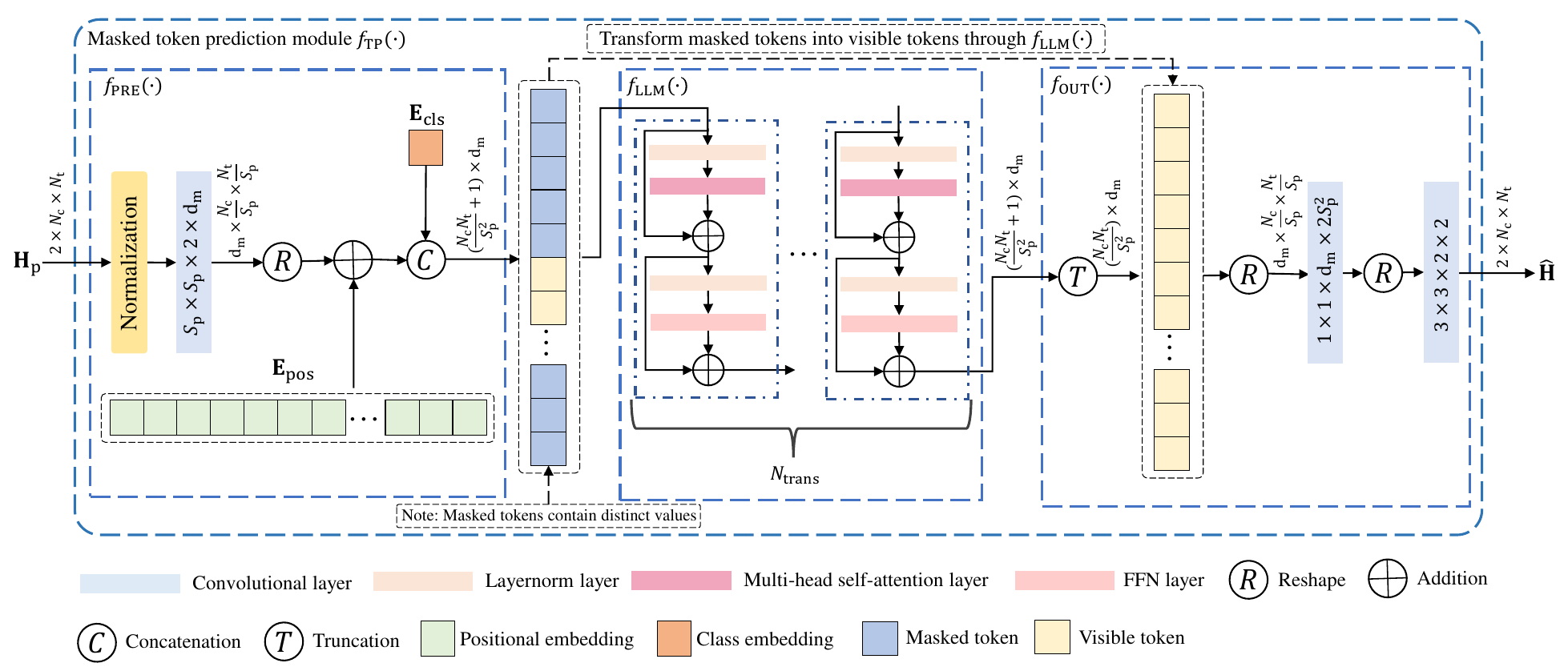}
	\caption{Design details of the masked token prediction module $f_{\mathrm{TP}}(\cdot)$.}
	\label{tpmodel}
\end{figure*}

The preliminary decoder $f_{\mathrm{PD}}(\cdot)$ is designed to assist the LLM in reconstructing the CSI by generating a coarse estimate of $\mathbf{H}$. Crucially, it provides a contextually informed, low-error initialization for the subsequent LLM-based prediction module, thereby significantly reducing the initial complexity of the masked token prediction task. The detailed network architecture is presented in Fig. \ref{fdecoder}.

$f_{\mathrm{PD}}(\cdot)$ first constructs a matrix $\mathbf{K} \in \mathbb{C}^{2 \times N_\mathrm{c} \times N_\mathrm{t}}$, where all elements are initialized to the mean value of $\mathbf{H}$. A fill operator is then applied to insert the values from $\mathbf{v}$ into their corresponding positions in $\mathbf{K}$, as specified by the indices in $\mathbf{p}$. The filled matrix is subsequently passed through two cascaded Res Networks followed by a convolutional layer to produce the preliminary reconstructed $\mathbf{H}_\mathrm{p}$. Each Res Network consists of three sequential convolutional layers and a residual connection. Specifically, the input to the Res Network is concatenated with the feature maps extracted by the three convolutional layers, and then passed through an activation function to obtain the final output of the Res Network. The residual connection helps mitigate the vanishing and exploding gradient problems during training, thereby improving model stability and convergence.

\subsection{Masked Token Prediction Module $f_{\mathrm{TP}}(\cdot)$}
The masked token prediction module $f_{\mathrm{TP}}(\cdot)$ serves as the core component underlying the performance improvement of the proposed LLMCsiNet. Its unique novelty lies in fundamentally reformulating the CSI problem by adapting the powerful sequence modeling of the pretrained LLM architecture through domain-specific innovations, including a physics-driven masking strategy and convolution-based tokenization. It transforms the preliminarily reconstructed CSI $\mathbf{H}_\mathrm{p}$ into a CSI token sequence comprising both masked and visible tokens, and subsequently leverages the LLM to predict the masked tokens based on the visible tokens, thereby finally refining CSI reconstruction. This module consists of three submodules: preprocessing NNs $f_{\mathrm{PRE}}(\cdot)$, pretrained LLM $f_{\mathrm{LLM}}(\cdot)$, and output NNs $f_{\mathrm{OUT}}(\cdot)$. The detailed architecture of $f_{\mathrm{TP}}(\cdot)$ is depicted in Fig. \ref{tpmodel}.

In $f_{\mathrm{PRE}}(\cdot)$, $\mathbf{H}_\mathrm{p}$ is first normalized to enhance training stability. We treat each CSI patch of size $S_\mathrm{p} \times S_\mathrm{p}$ as a CSI token, and the patch extraction is implemented by a convolutional layer. Specifically, the normalized $\mathbf{H}_\mathrm{p}$ is passed through a convolutional layer with the kernel size and stride both set to $S_\mathrm{p}$. The output has a channel dimension equal to $\mathrm{d}_\mathrm{m}$, which corresponds to the token embedding dimension required by the LLM. The resulting feature map is then reshaped into a token sequence of length $(\frac{N_\mathrm{c}N_\mathrm{t}}{S_\mathrm{p}^2})$, where each CSI token is a $\mathrm{d}_\mathrm{m}$-dimensional vector. Compared to directly partitioning CSI into non-overlapping patches, the use of convolution for patch extraction allows LLMCsiNet to adaptively learn optimal patch representations during training. Unlike convolutional neural networks, which leverage local receptive fields, or recurrent neural networks, which exploit sequential dependencies, transformers used as the backbone of LLMs lack inherent positional awareness. To compensate for this, we add positional embeddings $\mathbf{E}_\mathrm{pos} \in \mathbb{C}^{(\frac{N_\mathrm{c}N_\mathrm{t}}{S_\mathrm{p}^2}) \times \mathrm{d}_\mathrm{m}}$ to the CSI token sequence, encoding the position of each token in the CSI token sequence. In addition, a class embedding $\mathbf{E}_\mathrm{cls} \in \mathbb{C}^{1 \times \mathrm{d}_\mathrm{m}}$ is concatenated to the token sequence to aggregate global CSI information and assist the LLM in CSI reconstruction. As a result, the final input to the LLM is a token sequence of length $(\frac{N_\mathrm{c}N_\mathrm{t}}{S_\mathrm{p}^2}+1)$. 

In conclusion, through the processing of $f_{\mathrm{PRE}}(\cdot)$, the original CSI image is transformed into a CSI token sequence of length $(\frac{N_\mathrm{c}N_\mathrm{t}}{S_\mathrm{p}^2}+1)$. Among these tokens, a subset corresponds to high self-information CSI elements fed back from the UE to the BS and is regarded as visible tokens. The remaining tokens, corresponding to low self-information CSI elements not fed back from the UE to the BS, are regarded as masked tokens. It is important to note that the feature values corresponding to different masked tokens are distinct, as they have already undergone preliminary reconstruction by $f_{\mathrm{PD}}(\cdot)$. The CSI token sequence is subsequently fed into $f_{\mathrm{LLM}}(\cdot)$. $f_{\mathrm{LLM}}(\cdot)$ is responsible for predicting the masked tokens based on the visible tokens. $f_{\mathrm{LLM}}(\cdot)$ consists of $N_\mathrm{trans}$ transformer layers, which form the core architecture of the LLM. Each transformer layer comprises a multi-head self-attention layer, two layernorm layers, and a feed-forward neural network (FNN) layer. The multi-head self-attention layer captures long-range dependencies between tokens, enabling global context modeling across the sequence. The layernorm layer stabilizes and accelerates the training process by normalizing layer inputs, while the FNN layer introduces nonlinearity to enhance the expressive power of the LLM. Additionally, residual connections are incorporated within each transformer layer to mitigate issues related to vanishing and exploding gradients in deep networks. By adjusting $N_\mathrm{trans}$, a flexible trade-off can be achieved between network complexity and reconstruction performance to meet specific requirements in practice. Finally, the updated output sequence from $f_{\mathrm{LLM}}(\cdot)$, where masked tokens are predicted and thus become visible tokens, is passed to the output module $f_{\mathrm{OUT}}(\cdot)$. The primary function of $f_{\mathrm{OUT}}(\cdot)$ is to restore the sequence to the original CSI dimensions, thereby producing the final reconstructed CSI.

\subsection{Design of Training Procedure $f_{\mathrm{TRAIN}}(\cdot)$}
\begin{algorithm2e}[t]
\SetAlgoLined
\KwIn{Training dataset $\mathcal{T}$; Early stop patience $\mathcal{T}_{\text{patience}}$.}
\KwOut{Trained model parameters $\mathcal{O} = \{\mathcal{O}_{\mathrm{EN}}, \mathcal{O}_{\mathrm{PD}}, \mathcal{O}_{\mathrm{TP}}\}$.}

\textbf{Stage 1: Train $f_{\mathrm{EN}}(\cdot)$ and $f_{\mathrm{PD}}(\cdot)$ only} \;
Initialize $\mathcal{O}_{\mathrm{EN}}$, $\mathcal{O}_{\mathrm{PD}}$ randomly\;
Freeze parameters $\mathcal{O}_{\mathrm{TP}}$\;
\Repeat{the validation loss does not decrease for $\mathcal{T}_{\text{patience}}$ consecutive epochs}{
  \For{batch data $\mathbf{H}_i \in \mathcal{T}$}{
    $\mathbf{m}_i=[\mathbf{v}_i, \mathbf{p}_i] \gets f_{\mathrm{EN}}(\mathbf{H}_i; \mathcal{O}_{\mathrm{EN}})$\;
    $\mathbf{H}_\mathrm{p}^i \gets f_{\mathrm{PD}}(\mathbf{m}_i; \mathcal{O}_{\mathrm{PD}})$\;
    Compute loss $\mathcal{L}_1 = \|\widehat{\mathbf{H}}_\mathrm{p}^i - \mathbf{H}_i\|_2^2$\;
    Update $\mathcal{O}_{\mathrm{EN}}, \mathcal{O}_{\mathrm{PD}}$ using gradient descent\;
  }
}

\textbf{Stage 2: Jointly train all modules $f_{\mathrm{EN}}(\cdot)$, $f_{\mathrm{PD}}(\cdot)$, and $f_{\mathrm{TP}}(\cdot)$} \;
Unfreeze $\mathcal{O}_{\mathrm{TP}}$, freeze $\mathcal{O}_{\mathrm{EN}}$ and $\mathcal{O}_{\mathrm{PD}}$ if the network has difficulty converging\;
\Repeat{the validation loss does not decrease for $\mathcal{T}_{\text{patience}}$ consecutive epochs}{
  \For{batch data $\mathbf{H}_i \in \mathcal{T}$}{
    $\mathbf{m}_i=[\mathbf{v}_i, \mathbf{p}_i] \gets f_{\mathrm{EN}}(\mathbf{H}_i; \mathcal{O}_{\mathrm{EN}})$\;
    $\mathbf{H}_\mathrm{p}^i \gets f_{\mathrm{PD}}(\mathbf{m}_i; \mathcal{O}_{\mathrm{PD}})$\;
    $\widehat{\mathbf{H}}_i \gets f_{\mathrm{TP}}(\mathbf{H}_\mathrm{p}^i; \mathcal{O}_{\mathrm{TP}})$\;
    Compute loss $\mathcal{L}_2 = \|\widehat{\mathbf{H}}_i - \mathbf{H}_i\|_2^2$\;
    Update $\mathcal{O}_{\mathrm{EN}}$, $\mathcal{O}_{\mathrm{PD}}$, $\mathcal{O}_{\mathrm{TP}}$ using gradient descent\;
  }
}
\textbf{return} $\mathcal{O} = \{\mathcal{O}_{\mathrm{EN}}, \mathcal{O}_{\mathrm{PD}}, \mathcal{O}_{\mathrm{TP}}\}$

\caption{Two-Stage Training Procedure for LLMCsiNet}
\label{alg:llmcsi_training}
\end{algorithm2e}

In this section, we develop a training procedure, denoted as $f_{\mathrm{TRAIN}}(\cdot)$, for optimizing the proposed LLMCsiNet. The objective of $f_{\mathrm{TRAIN}}(\cdot)$ is to enable $f_{\mathrm{EN}}(\cdot)$ to effectively select and feed back the CSI elements with the highest self-information, while $f_{\mathrm{PD}}(\cdot)$ performs preliminary reconstruction of CSI. Finally, $f_{\mathrm{TP}}(\cdot)$ predicts the masked tokens based on the visible tokens to achieve high-accuracy CSI reconstruction. During the training process, we observe that jointly optimizing the pretrained LLM $f_{\mathrm{TP}}(\cdot)$ with randomly initialized modules $f_{\mathrm{EN}}(\cdot)$ and $f_{\mathrm{PD}}(\cdot)$ from the start leads to unstable optimization and slow convergence. In the early stage of joint training, $f_{\mathrm{EN}}(\cdot)$ has not yet learned to effectively extract CSI elements with high self-information. Simultaneously, $f_{\mathrm{PD}}(\cdot)$ produces preliminarily reconstructed CSI that contains considerable noise and distortion. If $f_{\mathrm{TP}}(\cdot)$, which incorporates a pretrained LLM, is fine-tuned at this stage, the LLM may learn incorrect prediction patterns based on low-quality CSI token inputs. 
\begin{table}[t]
    \center
    \caption{Channel Parameters}
    \setlength{\tabcolsep}{2mm} 
    \renewcommand{\arraystretch}{1.2} 
    \begin{tabular}{c|c|c}
    \Xhline{1.1 pt} Channel model &Parameter & Value\\
     \hline \multirow{10}*{COST2100out \cite{cost2100}}&Carrier frequency & $300$ MHz\\
     \cline{2-3}&Bandwidth & 20 MHz\\
     \cline{2-3}&Maximum delay spread & $5 \times 10^{-7}$ s\\
     \cline{2-3}&Number of paths & 48\\
     \cline{2-3}&Scenario & Outdoor, nLoS\\
     \cline{2-3}&$N_\mathrm{f}$ & $1024$ \\
     \cline{2-3}&$N_\mathrm{c}$ & $32$ \\
     \cline{2-3}&$N_\mathrm{t}$ & $32$ \\
     \hline \multirow{10}*{COST2100in \cite{cost2100}}&Carrier frequency & $5.3$ GHz\\
     \cline{2-3}&Bandwidth & 20 MHz \\
     \cline{2-3}&Maximum delay spread & $5 \times 10^{-7}$ s\\
     \cline{2-3}&Number of paths & 3\\
     \cline{2-3}&Scenario & Indoor, LoS\\
     \cline{2-3}&$N_\mathrm{f}$ & $1024$ \\
     \cline{2-3}&$N_\mathrm{c}$ & $32$ \\
     \cline{2-3}&$N_\mathrm{t}$ & $32$ \\
     \hline \multirow{10}*{UMa \cite{3gpp_38901}}&Carrier frequency & $28$ GHz\\
     \cline{2-3}&Bandwidth & 100 MHz\\
     \cline{2-3}&Maximum delay spread & $8 \times 10^{-6}$ s\\
     \cline{2-3}&Number of paths & 34\\
     \cline{2-3}&Scenario & Outdoor, nLoS\\
     \cline{2-3}&$N_\mathrm{f}$ & $1024$ \\
     \cline{2-3}&$N_\mathrm{c}$ & $32$ \\
     \cline{2-3}&$N_\mathrm{t}$ & $32$ \\
     \hline \multirow{10}*{DeepMIMOo1 \cite{deepmimo}}&Carrier frequency & $3.4$ GHz\\
     \cline{2-3}&Bandwidth & 10 MHz \\
     \cline{2-3}&Maximum delay spread & $1 \times 10^{-6}$ s\\
     \cline{2-3}&Number of paths & 10\\
     \cline{2-3}&Scenario & Outdoor, nLoS\\
     \cline{2-3}&$N_\mathrm{f}$ & $1024$ \\
     \cline{2-3}&$N_\mathrm{c}$ & $32$ \\
     \cline{2-3}&$N_\mathrm{t}$ & $32$ \\
    \Xhline{1.1pt}
    \end{tabular}
    \label{sim_para1}
\end{table}

To address this issue, we adopt a two-stage training strategy for the proposed LLMCsiNet. In the first stage, we train only  
$f_{\mathrm{EN}}(\cdot)$ and $f_{\mathrm{PD}}(\cdot)$ to provide stable and informative inputs for the LLM. In the second stage, we train the entire network by jointly training $f_{\mathrm{TP}}(\cdot)$ and the previously optimized $f_{\mathrm{EN}}(\cdot)$ and $f_{\mathrm{PD}}(\cdot)$. The decision of whether to freeze the parameters of $f_{\mathrm{EN}}(\cdot)$ and $f_{\mathrm{PD}}(\cdot)$ is made based on the convergence behavior of the network. If the network exhibits unstable or slow convergence, freezing parameters of $f_{\mathrm{EN}}(\cdot)$ and $f_{\mathrm{PD}}(\cdot)$ may help stabilize the training process. We additionally incorporate an early stopping mechanism to reduce training overhead and avoid overfitting. Specifically, training is terminated when the validation loss fails to improve for $\mathcal{T}_{\text{patience}}$ consecutive epochs, and the best-performing model is saved at the same time. The detailed training procedure is presented in Algorithm~\ref{alg:llmcsi_training}.

Let the overall trainable parameters of LLMCsiNet be denoted by $\mathcal{O} = \{\mathcal{O}_{\mathrm{EN}}, \mathcal{O}_{\mathrm{PD}}, \mathcal{O}_{\mathrm{TP}}\}$, where $\mathcal{O}_{\mathrm{EN}}$, $\mathcal{O}_{\mathrm{PD}}$, and $\mathcal{O}_{\mathrm{TP}}$ represent the parameters of $f_{\mathrm{EN}}(\cdot)$, $f_{\mathrm{PD}}(\cdot)$, and $f_{\mathrm{TP}}(\cdot)$, respectively. The complete training procedure for CSI reconstruction is described as 
\begin{align}
&\mathbf{H}_\mathrm{p} = f_{\mathrm{PD}}\left( f_{\mathrm{EN}}(\mathbf{H}; \mathcal{O}_{\mathrm{EN}}); \mathcal{O}_{\mathrm{PD}} \right) \label{eq:stage1}; \\
&\mathbf{\widehat{H}} =  
f_{\mathrm{TP}}\left( f_{\mathrm{PD}}\left( f_{\mathrm{EN}}(\mathbf{H}; \mathcal{O}_{\mathrm{EN}}); \mathcal{O}_{\mathrm{PD}} \right); \mathcal{O}_{\mathrm{TP}} \right). \label{eq:stage2}
\end{align}
We employ the mean squared error (MSE) as the loss function to optimize the network parameters via backpropagation. The loss function is defined as
\begin{equation}
\mathcal{L}(\mathcal{O}) = \frac{1}{D} \sum_{i=1}^{D} \left\| f_{\mathrm{TP}}\left( f_{\mathrm{PD}}\left( f_{\mathrm{EN}}(\mathbf{H}_{i}) \right) \right) - \mathbf{H}_{i} \right\|_2^2,
\end{equation}
where $D$ denotes the total number of training samples, while $\mathbf{H}_{i}$ represents the $i$-th original CSI sample. We adopt the normalized mean squared error (NMSE) and squared generalized cosine similarity (SGCS) as the evaluation metrics. The NMSE is defined as
\begin{equation}
\mathrm{NMSE} = \mathbb{E} \left[ \frac{\| \mathbf{H}_{i} - \mathbf{\widehat{H}}_{i} \|_2^2}{\| \mathbf{H}_{i} \|_2^2} \right],
\end{equation}
where $\mathbf{\widehat{H}}_{i}$ represents the corresponding reconstructed sample of the $i$-th original CSI sample, $\mathbf{H}_{i}$. The SGCS is defined as
\begin{equation}
\mathrm{SGCS} = \mathbb{E} \left[ \frac{|\mathrm{Tr}(\mathbf{H}_{i}^{\mathrm{H}}\mathbf{\widehat{H}}_{i})|^2}{\|\mathbf{H}_{i}\|_F^2 \|\mathbf{\widehat{H}}_{i}\|_F^2} \right].
\end{equation}

In terms of practical operation, the framework is governed by a distinct two-phase process: the LLMCsiNet is first fully optimized via an offline training stage, followed by strategic online deployment. The final deployment utilizes an architectural split, where the lightweight encoding network operates at the UE, and the resource-intensive LLM-based decoding network is hosted exclusively at the BS, ensuring efficient real-time performance.

\section{Experimental Results}
\begin{figure*}[t]
\centering
\subfigure[$\sigma=1/8$]{
\includegraphics[width=8.5cm]{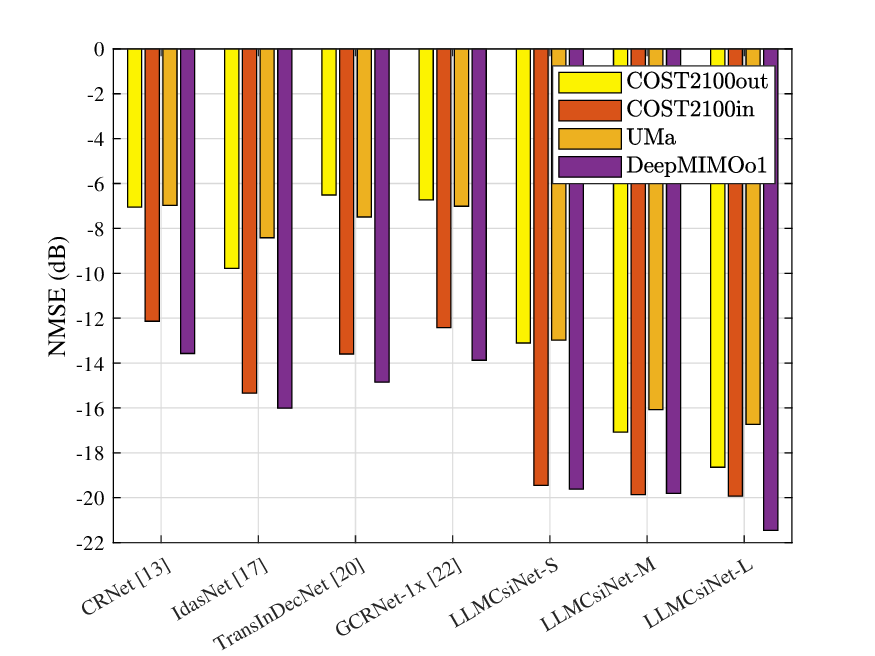}
\label{Fig3-1}
}
\quad
\subfigure[$\sigma=1/16$]{
\includegraphics[width=8.5cm]{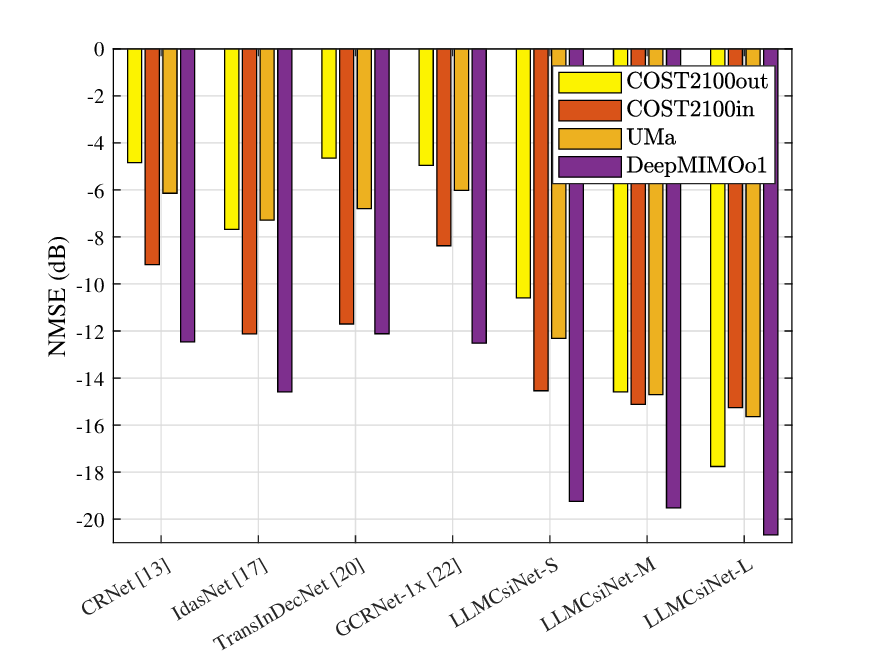}
\label{Fig3-2}
}
\quad
\subfigure[$\sigma=1/32$]{
\includegraphics[width=8.5cm]{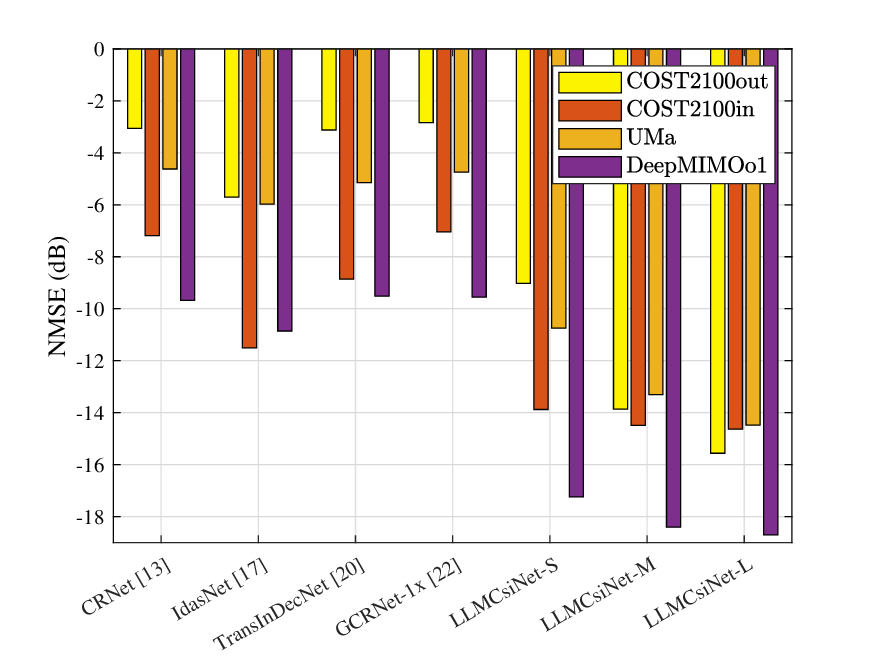}
\label{Fig3-3}
}
\quad
\subfigure[$\sigma=1/64$]{
\includegraphics[width=8.5cm]{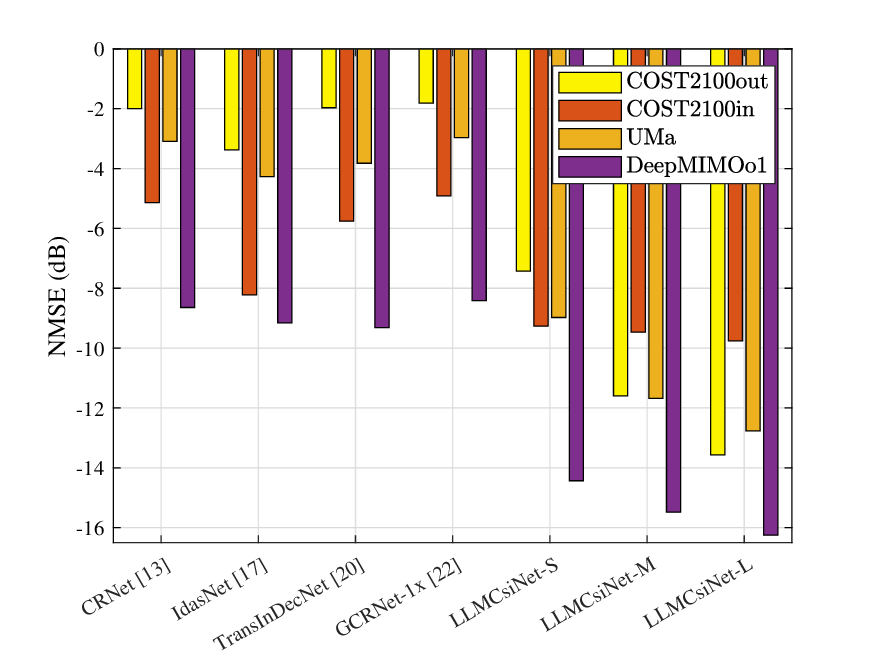}
\label{Fig3-4}
}
\caption{NMSE comparison under different choices of compression ratio $\sigma$.}\label{Fig3}
\end{figure*}

In this section, we conduct a comprehensive series of simulations to evaluate the performance advantages of the proposed LLMCsiNet. We begin by presenting the detailed simulation settings. Subsequently, we compare the NMSE and SGCS performance of LLMCsiNet with small models under different channel environments and compression ratios. We then explore the NMSE performance of LLMCsiNet in challenging scenarios with multiple compression ratios and moving UEs. In addition, we conduct ablation studies to evaluate the impact of each module on reconstruction accuracy. We also evaluate the robustness of LLMCsiNet in cross-scenario transfer learning. Finally, we evaluate the multi-user communication rate improvements offered by LLMCsiNet. 
\subsection{Simulation Setup and Training Procedure}
We evaluate our proposed method using four distinct channel datasets: COST2100out, COST2100in, UMa, and DeepMIMOo1. The COST2100out and COST2100in datasets are generated based on the COST2100 channel model \cite{cost2100}, representing indoor LoS and outdoor NLoS scenarios, respectively. The UMa dataset is meticulously generated according to the 3GPP TR 38.901 standard \cite{3gpp_38901}, while the DeepMIMOo1 dataset is generated using the DeepMIMO software \cite{deepmimo}, specifically utilizing the Outdoor1 channel environment. The inclusion of these datasets, which introduce complexities related to mmWave propagation and ray-tracing models, further ensures comprehensive evaluation. For consistency across all four environments, the size of the training set is uniformly set to $20,000$ samples, while the validation and testing sets each contain $10,000$ samples. The detailed channel parameters utilized for all four datasets are summarized in Table~\ref{sim_para1}.

In the first training stage, all weights and biases of the fully-connected and convolutional layers in both $f_{\mathrm{EN}}(\cdot)$ and $f_{\mathrm{PD}}(\cdot)$ are randomly initialized. In the second training stage, $f_{\mathrm{EN}}(\cdot)$ and $f_{\mathrm{PD}}(\cdot)$ load the trained parameters obtained from the first training stage. For $f_{\mathrm{TP}}(\cdot)$, the $f_{\mathrm{LLM}}(\cdot)$ is initialized with the pretrained parameters of a pretrained LLM, while the remaining submodules are also randomly initialized. In particular, we adopt GPT-2 Large \cite{gpt2} as the pretrained LLM in this paper. Pretrained decoder-only models such as LLaMA \cite{llmsel2-2} and Gemma3 \cite{llmsel2-3} are also highly suitable alternatives due to their shared transformer architecture. The Manhattan radius $R$ is set to $3$. The size of CSI patch $S_\mathrm{p}$ is set to $4$. The Adam optimizer is employed for parameter optimization. Both the first and second training stages are trained for $300$ epochs, with the early stopping patience parameter $\mathcal{T}_{\text{patience}}$ set to $20$. All simulations are implemented using PyTorch on two NVIDIA RTX 4090 GPUs. A cosine annealing learning rate schedule \cite{deepcsi2} is applied, where the learning rate first increases from 0 to $2 \times 10^{-3}$ and then decays according to 
\begin{equation}
    \gamma=\gamma_\mathrm{min}+\frac{1}{2}(\gamma_\mathrm{max}-\gamma_\mathrm{min})\left(1+\mathrm{\cos}\left(\frac{t-T_\mathrm{w}}{T_\mathrm{t}-T_\mathrm{w}}\pi\right)\right),
\end{equation}
where $t$ denotes the current epoch index, while $T_\mathrm{w}$ and $T_\mathrm{t}$ denote the number of epochs during the warm-up stage and the total number of training epochs, respectively. The learning rate at epoch $t$ is denoted by $\gamma$, with $\gamma_\mathrm{max}$ and $\gamma_\mathrm{min}$ representing the initial and final learning rates, respectively.

\subsection{Performance of the Proposed LLMCsiNet}

\begin{table*}[ht]
    \center
    \caption{A Balanced Comparison of Transmitting Bits under Different $\sigma$}
    \setlength{\tabcolsep}{4mm} 
    \renewcommand{\arraystretch}{1} 
    \begin{tabular}{c|c|cc|cc|c}
    \Xhline{1.1pt} \multirow{2}*{Methods} & Compression & Number of & Transmitting & Number of  &Transmitting & \multirow{2}*{Total bits}\\
    &ratio $\sigma$&codeword values &bits& position index&bits&\\
    \hline {CRNet \cite{deepcsi2}} & \multirow{2}*{1/8} & 256 & 64 & 0 & 0 & 16,384\\
    {\textbf{LLMCsiNet}} & & 221+\textbf{1} & 64 & 221 & 10 & \textbf{16,418}\\
    \hline {CRNet \cite{deepcsi2}} & \multirow{2}*{1/16} & 128 & 64 & 0 & 0 & 8,192\\
    {\textbf{LLMCsiNet}} & & 111+\textbf{1} & 64 & 111 & 10 & \textbf{8,278}\\
    \hline {CRNet \cite{deepcsi2}} & \multirow{2}*{1/32} & 64 & 64 & 0 & 0 & 4,096\\
    {\textbf{LLMCsiNet}} & & 56+\textbf{1} & 64 & 56 & 10 & \textbf{4,208}\\
    \hline {CRNet \cite{deepcsi2}} & \multirow{2}*{1/64} & 32 & 64 & 0 & 0 & 2,048\\
    {\textbf{LLMCsiNet}} & & 28+\textbf{1} & 64 & 28 & 10 & \textbf{2,136}\\
    \Xhline{1.1pt}
    \end{tabular}
    \label{compressionratio}
\end{table*}

\begin{table}[ht]
    \center
   \caption{SGCS Comparison under Different Scenarios}
    \setlength{\tabcolsep}{2mm} 
    \renewcommand{\arraystretch}{1} 
    \scalebox{0.8}{
    \begin{tabular}{c|c|c|c|c|c}
    \Xhline{1.1pt}  {Methods} & {$\sigma$} & COST2100out & COST2100in & UMa & DeepMIMOo1 \\
     \hline{CRNet \cite{deepcsi2}} &\multirow{7}*{1/8}&0.802&0.933&0.793&0.953 \\
     {IdasNet \cite{deepcsi6}} &&0.897&0.965 &0.850&0.971\\
     {TransInDecNet \cite{TransInDecNet}} &&0.776&0.949&0.824&0.963 \\
     {GCRNet-1x \cite{lightcsi2}} &&0.793&0.933&0.805&0.955 \\
    \textbf{LLMCsiNet-S} && \textbf{0.952} &\textbf{0.987} &\textbf{0.947}&\textbf{0.989}\\
    \textbf{LLMCsiNet-M} && \textbf{0.980} &\textbf{0.989}&\textbf{0.969}&\textbf{0.992}\\
    \textbf{LLMCsiNet-L} && \textbf{0.985} &\textbf{0.991}&\textbf{0.972}&\textbf{0.993}\\
    \hline{CRNet \cite{deepcsi2}} &\multirow{7}*{1/16}&0.671&0.874&0.750&0.946\\
     {IdasNet \cite{deepcsi6}} &&0.808&0.933&0.808&0.959\\
     {TransInDecNet \cite{TransInDecNet}} &&0.659&0.925&0.777&0.943\\
     {GCRNet-1x \cite{lightcsi2}} &&0.672&0.864&0.743&0.944 \\
    \textbf{LLMCsiNet-S} && \textbf{0.914} &\textbf{0.963}&\textbf{0.938}&\textbf{0.987}\\
    \textbf{LLMCsiNet-M} && \textbf{0.968} &\textbf{0.968}&\textbf{0.961}&\textbf{0.988}\\
    \textbf{LLMCsiNet-L} && \textbf{0.982} &\textbf{0.969}&\textbf{0.967}&\textbf{0.991}\\
    \hline{CRNet \cite{deepcsi2}} &\multirow{7}*{1/32}&0.509&0.805&0.680&0.906 \\
     {IdasNet \cite{deepcsi6}} &&0.731&0.925&0.744&0.933\\
     {TransInDecNet \cite{TransInDecNet}} &&0.522&0.853& 0.706&0.904\\
     {GCRNet-1x \cite{lightcsi2}} &&0.504&0.786&0.691&0.901 \\
    \textbf{LLMCsiNet-S} && \textbf{0.876} &\textbf{0.957}&\textbf{0.912}&\textbf{0.981}\\
    \textbf{LLMCsiNet-M} && \textbf{0.959} &\textbf{0.963}&\textbf{0.947}&\textbf{0.986}\\
    \textbf{LLMCsiNet-L} && \textbf{0.974} &\textbf{0.965}&\textbf{0.955}&\textbf{0.987}\\
    \hline{CRNet \cite{deepcsi2}} &\multirow{7}*{1/64}&0.379&0.694&0.509&0.873\\
     {IdasNet \cite{deepcsi6}} &&0.549&0.846&0.631&0.890\\
     {TransInDecNet \cite{TransInDecNet}} &&0.378&0.725&0.595&0.901 \\
     {GCRNet-1x \cite{lightcsi2}} &&0.374&0.685&0.506&0.870 \\
    \textbf{LLMCsiNet-S} && \textbf{0.821} &\textbf{0.881}&\textbf{0.868}&\textbf{0.970}\\
    \textbf{LLMCsiNet-M} && \textbf{0.934} &\textbf{0.888}&\textbf{0.918}&\textbf{0.977}\\
    \textbf{LLMCsiNet-L} && \textbf{0.958} &\textbf{0.896}&\textbf{0.928}&\textbf{0.981}\\
    \Xhline{1.1pt}
    \end{tabular}}
    \label{SGCSperformance}
\end{table}

\begin{table*}[ht]
    \center
   \caption{Network Complexity Comparison}
    \setlength{\tabcolsep}{1mm} 
    \renewcommand{\arraystretch}{1.2} 
    \begin{tabular}{c|cc|cc|cc|cc}
    \Xhline{1.1pt}  \multirow{3}*{Methods} & \multicolumn{8}{c}{Compression ratio $\sigma$}\\
     \cline{2-9}&\multicolumn{2}{c|}{$\sigma$=1/8}&\multicolumn{2}{c|}{$\sigma$=1/16}&\multicolumn{2}{c|}{$\sigma$=1/32}&\multicolumn{2}{c}{$\sigma$=1/64}\\
     \cline{2-9}&Modules at UE&Modules at BS&Modules at UE&Modules at BS&Modules at UE&Modules at BS&Modules at UE&Modules at BS\\
    \hline {CRNet \cite{deepcsi2}} &0.52M & 0.53M& 0.26M & 0.27M&0.13M & 0.14M& 0.07M & 0.07M\\
    \hline {IdasNet \cite{deepcsi6}} &3.16K & 3.36K& 3.16K & 3.36K&3.16K & 3.36K&3.16K & 3.36K\\
    \hline {TransInDecNet \cite{TransInDecNet}} &0.80M & 0.52M &0.26M &0.54M &0.13M & 0.41M &0.07M & 0.34M \\
    \hline {GCRNet-1x \cite{lightcsi2}} & 0.52M &2.18M & 0.26M & 1.92M &0.13M & 1.79M &0.07M & 1.72M\\
    \hline \textbf{LLMCsiNet-S} &\textbf{3.16K} & 19.85M& \textbf{3.16K} & 19.85M&\textbf{3.16K} & 19.85M&\textbf{3.16K} & 19.85M\\
    \hline \textbf{LLMCsiNet-M} &\textbf{3.16K} & 59.21M& \textbf{3.16K} & 59.21M&\textbf{3.16K} & 59.21M&\textbf{3.16K} & 59.21M\\
    \hline \textbf{LLMCsiNet-L} &\textbf{3.16K} & 118.24M& \textbf{3.16K} & 118.24M&\textbf{3.16K} & 118.24M&\textbf{3.16K} & 118.24M\\
    \Xhline{1.1pt}
    \end{tabular}
    \label{networkcomplex}
\end{table*}

\begin{table*}[ht]
    \caption{Computational Complexity Comparison}
    \centering
    \setlength{\tabcolsep}{3pt} 
    \renewcommand{\arraystretch}{1.2} 
    \scalebox{0.95}{
    \begin{tabular}{c|c|c|c}
    \Xhline{1.1pt} \multicolumn{2}{c|}{Methods} &FLOPs&Big-$O$ Notation\\
    \hline \multirow{2}*{CRNet \cite{deepcsi2}}&Modules at UE &$152 N_\mathrm{c} N_\mathrm{t}+4 \sigma N_\mathrm{c}^2 N_\mathrm{t}^2$ & $O(N_\mathrm{c} N_\mathrm{t}+\sigma N_\mathrm{c}^2 N_\mathrm{t}^2)$\\
    \cline{2-4} &Modules at BS&$2802 N_\mathrm{c} N_\mathrm{t}+4 \sigma N_\mathrm{c}^2 N_\mathrm{t}^2$&$O(N_\mathrm{c} N_\mathrm{t}+\sigma N_\mathrm{c}^2 N_\mathrm{t}^2)$\\
    \hline \multirow{2}*{IdasNet \cite{deepcsi6}} &Modules at UE&$2313N_\mathrm{c} N_\mathrm{t}+\log(2 N_\mathrm{c} N_\mathrm{t})+64 N_\mathrm{c} N_\mathrm{t}\log(N_\mathrm{c} N_\mathrm{t})$ & $O(N_\mathrm{c} N_\mathrm{t}+\log(N_\mathrm{c} N_\mathrm{t})+N_\mathrm{c} N_\mathrm{t}\log(N_\mathrm{c} N_\mathrm{t}))$\\
    \cline{2-4} &Modules at BS&$3204 N_\mathrm{c} N_\mathrm{t}$& $O(N_\mathrm{c} N_\mathrm{t})$\\
    \hline \multirow{2}*{TransInDecNet \cite{TransInDecNet}} &Modules at UE&$4 \sigma N_\mathrm{c}^2 N_\mathrm{t}^2$ & $O(\sigma N_\mathrm{c}^2 N_\mathrm{t}^2)$\\
    \cline{2-4} &Modules at BS&$4 \sigma N_\mathrm{c}^2 N_\mathrm{t}^2+24N_\mathrm{c}^2\mathrm{d}_\mathrm{m}+20N_\mathrm{c}\mathrm{d}_\mathrm{m}^2+8N_\mathrm{c}\mathrm{d}_\mathrm{m}\mathrm{d}_\mathrm{ff}$& $O(\sigma N_\mathrm{c}^2 N_\mathrm{t}^2+N_\mathrm{c}^2\mathrm{d}_\mathrm{m}+N_\mathrm{c}\mathrm{d}_\mathrm{m}^2+N_\mathrm{c}\mathrm{d}_\mathrm{m}\mathrm{d}_\mathrm{ff})$\\
    \hline \multirow{2}*{GCRNet-1x \cite{lightcsi2}}&Modules at UE &$188 N_\mathrm{c} N_\mathrm{t}+4 \sigma N_\mathrm{c}^2 N_\mathrm{t}^2$ & $O(N_\mathrm{c} N_\mathrm{t}+\sigma N_\mathrm{c}^2 N_\mathrm{t}^2)$\\
    \cline{2-4} &Modules at BS&$1780 N_\mathrm{c} N_\mathrm{t}+4 \sigma N_\mathrm{c}^2 N_\mathrm{t}^2$&$O(N_\mathrm{c} N_\mathrm{t}+\sigma N_\mathrm{c}^2 N_\mathrm{t}^2)$\\
    \hline \multirow{2}*{LLMCsiNet} &Modules at UE&$2313N_\mathrm{c} N_\mathrm{t}+\log(2 N_\mathrm{c} N_\mathrm{t})+64 N_\mathrm{c} N_\mathrm{t}\log(N_\mathrm{c} N_\mathrm{t})$ & $O(N_\mathrm{c} N_\mathrm{t}+\log(N_\mathrm{c} N_\mathrm{t})+N_\mathrm{c} N_\mathrm{t}\log(N_\mathrm{c} N_\mathrm{t}))$ \\
    \cline{2-4} &Modules at BS&$3204 N_\mathrm{c} N_\mathrm{t}+2\mathrm{d}_\mathrm{m} N_\mathrm{c} N_\mathrm{t}+N_{\mathrm{trans}}(4L\mathrm{d}_\mathrm{m}^2+2L^2\mathrm{d}_\mathrm{m}+2 \mathrm{d}_\mathrm{m} \mathrm{d}_\mathrm{ff})$& $O(N_\mathrm{c} N_\mathrm{t}+\mathrm{d}_\mathrm{m} N_\mathrm{c} N_\mathrm{t}+N_{\mathrm{trans}}\mathrm{d}_\mathrm{m}(L\mathrm{d}_\mathrm{m}+L^2+2 \mathrm{d}_\mathrm{ff}))$\\
    \Xhline{1.1pt}
    \end{tabular}}
    \vspace{1ex} 
    
    \begin{minipage}{\linewidth} 
    \small
    In the table above, $N_\mathrm{c}$ is the number of retained non-zero rows in the angle-delay domain; $N_\mathrm{t}$ is the number of antennas; $\sigma$ represents the compression ratio. For the LLM-related parameters: $N_{\mathrm{trans}}$ denotes the number of transformer layers; $L=(\frac{N_\mathrm{c}N_\mathrm{t}}{S_\mathrm{p}^2}+1)$ is the token sequence length with patch size $S_\mathrm{p}$; $\mathrm{d}_\mathrm{m}$ is the model dimension; and $\mathrm{d}_\mathrm{ff}$ is the feed-forward network dimension.
    \end{minipage}
    \label{compucomplex1}
\end{table*}

\begin{table}[ht]
    \center
   \caption{NMSE Comparison with Quantization under Different Scenarios}
    \setlength{\tabcolsep}{2mm} 
    \renewcommand{\arraystretch}{1} 
    \begin{tabular}{c|c|c|c}
    \Xhline{1.1pt}  {Methods} & {$\sigma$} & COST2100out & COST2100in\\
     \hline{CRNet \cite{deepcsi2}} &\multirow{7}*{1/8}&-6.932&-11.729 \\
     {IdasNet \cite{deepcsi6}} && -9.638&-14.637\\
    \textbf{LLMCsiNet-S} && \textbf{-13.032} &\textbf{-18.273}\\
    \textbf{LLMCsiNet-M} && \textbf{-16.841} &\textbf{-18.869}\\
    \textbf{LLMCsiNet-L} && \textbf{-18.194} &\textbf{-18.961}\\
    \hline{CRNet \cite{deepcsi2}} &\multirow{7}*{1/16}&-4.739&-8.917 \\
     {IdasNet \cite{deepcsi6}} && -7.072&-12.018\\
    \textbf{LLMCsiNet-S} && \textbf{-10.553} &\textbf{-14.452}\\
    \textbf{LLMCsiNet-M} && \textbf{-14.538} &\textbf{-15.025}\\
    \textbf{LLMCsiNet-L} && \textbf{-17.677} &\textbf{-15.078}\\
    \hline{CRNet \cite{deepcsi2}} &\multirow{7}*{1/32}&-3.042&-7.118 \\
     {IdasNet \cite{deepcsi6}} && -5.688&-11.227\\
    \textbf{LLMCsiNet-S} && \textbf{-9.012} &\textbf{-13.835}\\
    \textbf{LLMCsiNet-M} && \textbf{-13.796} &\textbf{-14.435}\\
    \textbf{LLMCsiNet-L} && \textbf{-15.434} &\textbf{-14.569}\\
    \hline{CRNet \cite{deepcsi2}} &\multirow{7}*{1/64}&-1.986&-5.119\\
     {IdasNet \cite{deepcsi6}} && -3.358&-8.208\\
    \textbf{LLMCsiNet-S} && \textbf{-7.335} &\textbf{-9.250}\\
    \textbf{LLMCsiNet-M} && \textbf{-11.458} &\textbf{-9.443}\\
    \textbf{LLMCsiNet-L} && \textbf{-13.496} &\textbf{-9.624}\\
    \Xhline{1.1pt}
    \end{tabular}
    \label{Quantizationperformance}
\end{table}

\begin{figure}[t]
	\centering  
	\includegraphics[width=8.5cm]{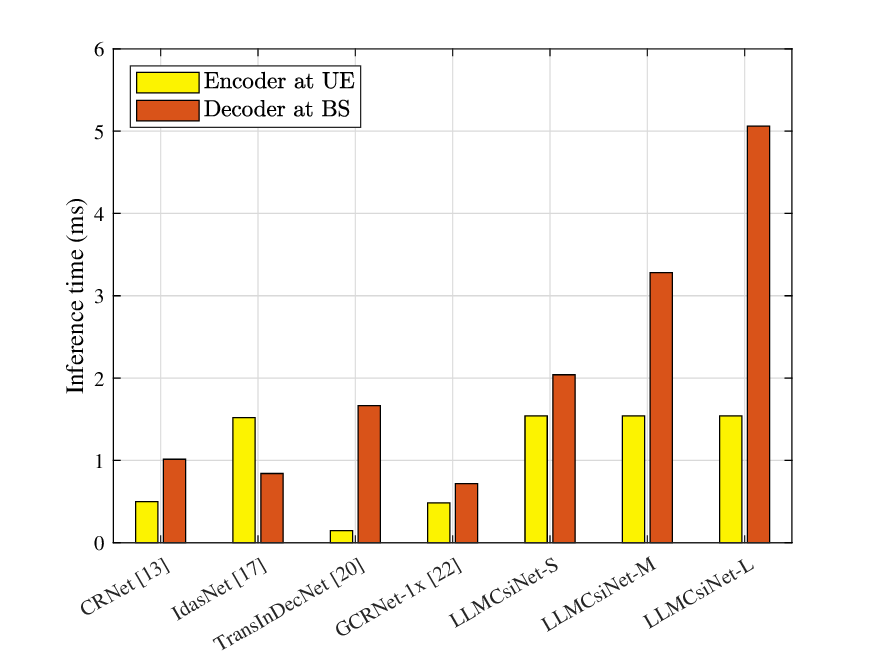}
	\caption{Comparison of inference time of the encoder at UE and the decoder at BS.}
	\label{infertime1}
\end{figure}

To demonstrate the effectiveness of the proposed LLMCsiNet for CSI feedback, we compare its performance against several state-of-the-art baselines under various compression ratios. Specifically, simulations are conducted at compression ratios $\sigma$ of $1/8$, $1/16$, $1/32$, and $1/64$. We utilize NMSE and SGCS as key evaluation metrics, alongside the complexity.

Since the index vector $\mathbf{p}$ is additionally fed back, its transmission overhead must be considered in the compression ratio calculation. The compression ratio $\sigma$ of LLMCsiNet is defined as
\begin{equation}
\sigma = \frac{q_1 \times (M+1) + q_2 \times M}{q_1 \times 2 \times N_c \times N_t},
\end{equation}
where $M$ denotes the number of CSI elements that are selected for feedback, $q_1$ represents the number of bits required to transmit each codeword value, and $q_2$ denotes the number of bits required to transmit each position index. The comparison of transmitting bits under different $\sigma$ is presented in Table \ref{compressionratio}.

We consider four state-of-the-art baselines, CRNet \cite{deepcsi2}, IdasNet \cite{deepcsi6}, TransInDecNet \cite{TransInDecNet}, and GCRNet-1x \cite{lightcsi2}. The proposed LLMCsiNet is implemented in three variants, namely LLMCsiNet-S, LLMCsiNet-M, and LLMCsiNet-L, which differ in the number of LLM transformer layers $N_{\mathrm{trans}}$ used in the $f_{\mathrm{LLM}}(\cdot)$ module. In LLMCsiNet-S, $N_{\mathrm{trans}}$ is set to 1; in LLMCsiNet-M, it is set to $3$; and in LLMCsiNet-L, it is set to $6$. The comprehensive simulation results, spanning all four channel datasets, are presented in Fig. \ref{Fig3} (for NMSE) and Table \ref{SGCSperformance} (for SGCS).

\subsubsection{NMSE Performance}

In the COST2100out channels, which represent a complex outdoor nLoS environment featuring rich multipath and nonlinear fading characteristics, we do not freeze the parameters of $f_{\mathrm{EN}}(\cdot)$ and $f_{\mathrm{PD}}(\cdot)$ during the second stage of joint training. Under a compression ratio of $1/8$, the NMSE values achieved by the baseline methods are: CRNet ($-7.051$ dB), IdasNet ($-9.783$ dB), TransInDecNet ($-6.512$ dB), and GCRNet-1x ($-6.727$ dB). In stark contrast, the proposed LLMCsiNet-S, LLMCsiNet-M, and LLMCsiNet-L achieve $-13.107$ dB, $-17.078$ dB, and $-18.643$ dB, respectively. These results indicate that LLMCsiNet provides a notable performance gain ranging from approximately $3.5$ dB to $9$ dB over all baselines as the LLM model capacity increases. This significant advantage is maintained even in extreme compression regimes. When the compression ratio $\sigma$ is further reduced to $1/64$, LLMCsiNet-L still achieves $-13.566$ dB, far surpassing the best baseline performance. Due to limited network size, conventional small models and recent lightweight variants struggle to extract and compress the high-dimensional information effectively, highlighting the necessity of the LLM’s large capacity.

In the COST2100in channels, we freeze the parameters of $f_{\mathrm{EN}}(\cdot)$ and $f_{\mathrm{PD}}(\cdot)$ during the second stage of joint training. Under the compression ratio of $1/8$, the baselines achieve NMSE performances ranging from $-12.138$ dB (CRNet) to $-15.342$ dB (IdasNet). In comparison, LLMCsiNet-S, LLMCsiNet-M, and LLMCsiNet-L achieve NMSE performances of $-19.449$ dB, $-19.865$ dB, and $-19.926$ dB, respectively. These results indicate that LLMCsiNet offers a performance improvement of approximately $5$ dB over conventional models. However, in contrast to the COST2100out results, the performance gain from increasing the LLM model scale is relatively limited. This is primarily because COST2100in channels exhibit LoS characteristics, leading to sparser CSI distributions. As the indoor CSI is easier to represent and reconstruct, even small models can effectively recover the essential features, thereby diminishing the differential advantage of LLM scaling. Nevertheless, LLMCsiNet still demonstrates significant performance advantages over all four conventional baseline models across all compression ratios, consistently achieving NMSE gains of $2$ to $3$ dB.

In the UMa channels, which represent a critical high-frequency mmWave outdoor environment, we do not freeze the parameters of $f_{\mathrm{EN}}(\cdot)$ and $f_{\mathrm{PD}}(\cdot)$ during joint training. The NMSE results at $\sigma=1/8$ show that all four baseline methods are significantly outperformed: CRNet achieves $-6.971$ dB, IdasNet reaches $-8.416$ dB, and the advanced TransInDecNet achieves $-7.492$ dB. In sharp contrast, LLMCsiNet-L achieves an NMSE of $-16.732$ dB, demonstrating a substantial performance gain of over $9$ dB compared to the strongest baseline. Furthermore, LLMCsiNet variants consistently maintain impressive NMSE gains, ranging from $5$ dB to $10$ dB, across all compression ratios over all four baseline methods. This confirms the architectural effectiveness of the LLM in modeling the sparse scattering characteristics and complex non-linearities inherent in mmWave channels, thereby ensuring high-fidelity reconstruction where conventional approaches struggle.

In the DeepMIMOo1 channels, which are generated via ray-tracing and require the precise recovery of deterministic geometric features, we freeze the parameters of $f_{\mathrm{EN}}(\cdot)$ and $f_{\mathrm{PD}}(\cdot)$ during joint training to focus the learning burden on the LLM. At $\sigma=1/8$, LLMCsiNet-L attains the highest NMSE score of $-21.456$ dB. This performance exceeds the strongest baseline, IdasNet ($-16.006$ dB), by more than $5$ dB and significantly surpasses TransInDecNet ($-14.851$ dB). The LLM's superior predictive capacity is evident even at the extreme compression ratio of $\sigma=1/64$, where LLMCsiNet-L ($-16.250$ dB) maintains robust reconstruction accuracy. This resilience, especially in low-bitrate regimes, conclusively proves the LLM's superior capability in capturing and reconstructing the complex, high-fidelity channel features determined by the deterministic ray-tracing process.

\subsubsection{SGCS Performance}
To assess the impact of reconstruction accuracy on precoding efficiency, we analyze the SGCS performance, the metric standardized by 3GPP. As shown in Table \ref{SGCSperformance}, the SGCS results strongly corroborate the NMSE findings, demonstrating that LLMCsiNet not only minimizes reconstruction error but also maximizes spatial correlation accuracy. LLMCsiNet variants consistently achieve the highest SGCS scores across all four channel environments and compression ratios. At $\sigma=1/8$, LLMCsiNet-L achieves near-perfect SGCS scores. The superiority is most striking in low-bitrate regimes; at $\sigma=1/64$, while baselines like CRNet degrade significantly, LLMCsiNet-L maintains high SGCS values, confirming its exceptional ability to recover the essential spatial direction information with high fidelity.

\subsubsection{Complexity Analysis}
We conduct a comprehensive complexity analysis encompassing parameter count, theoretical computational cost, and practical inference latency.

 As shown in Table \ref{networkcomplex} and Table \ref{compucomplex1}, LLMCsiNet features an intentionally asymmetric design. The model maintains an exceptionally low complexity at the UE side, with both parameter count and theoretical FLOPs matching the efficient IdasNet baseline. This ensures zero additional computational burden on the mobile terminal. While the BS complexity is necessarily higher due to the integrated LLM, this increase is justified by the superior performance and is aligned with our asymmetric philosophy of leveraging the BS's abundant resources. The high parameter count at the BS is further mitigated by the provision of scalable model variants (LLMCsiNet-S) and the applicability of general LLM compression techniques.

Additionally, the evaluation in Fig. \ref{infertime1} confirmed the method's real-time feasibility. At the delay-sensitive UE side, LLMCsiNet requires only $1.54$ ms, comparable to lightweight baselines like IdasNet. For the BS decoder, even the largest variant (LLMCsiNet-L) requires just $5.06$ ms. This efficiency stems from the highly parallelizable nature of transformer architectures, which effectively utilize GPU resources, demonstrating that the practical latency is sufficiently low for real-time CSI feedback.

\subsection{NMSE Comparison with Quantization Feedback}
In practical communication scenarios, continuous compressed codewords must be quantized before being fed back to BS via a limited transmission link. Consequently, we conduct simulations to evaluate the NMSE performance of CRNet \cite{deepcsi2}, IdasNet \cite{deepcsi6}, and three variants of LLMCsiNet under quantization. The Lloyd-Max algorithm is employed for quantizing the compressed codewords, and the simulation results are presented in Table \ref{Quantizationperformance}. The proposed LLMCsiNet demonstrates superior NMSE performance across different channels and compression ratios compared to small models, highlighting the effectiveness of LLMs in real-world scenarios.

\subsection{NMSE Comparison with Noisy CSI}
\begin{figure}[t]
	\centering 
	\includegraphics[width=8.5cm]{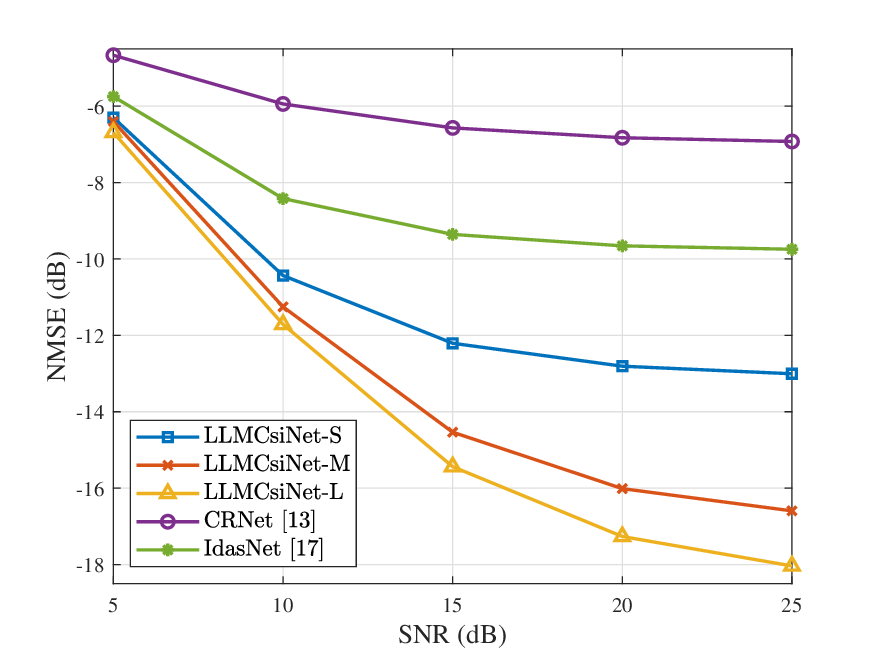}
	\caption{NMSE comparison with noisy CSI in the COST2100out channels ($\sigma=1/8$).}
	\label{noisenmse}
\end{figure}
In practical scenarios, CSI estimated by UE is affected by noise. To address this, we conduct simulations to evaluate the NMSE performance with noisy CSI feedback. Specifically, we introduce Gaussian noise at various dB levels into the ideal CSI input to the encoder and conduct simulations in the COST2100out channel environment with $\sigma=1/8$.

The results presented in Fig. \ref{noisenmse} indicate that even with noisy CSI inputs, the proposed LLMCsiNet achieves superior NMSE reconstruction accuracy, significantly outperforming small models. This demonstrates the excellent robustness of the proposed LLMCsiNet.

\begin{figure}[t]
	\centering 
	\includegraphics[width=8.5cm]{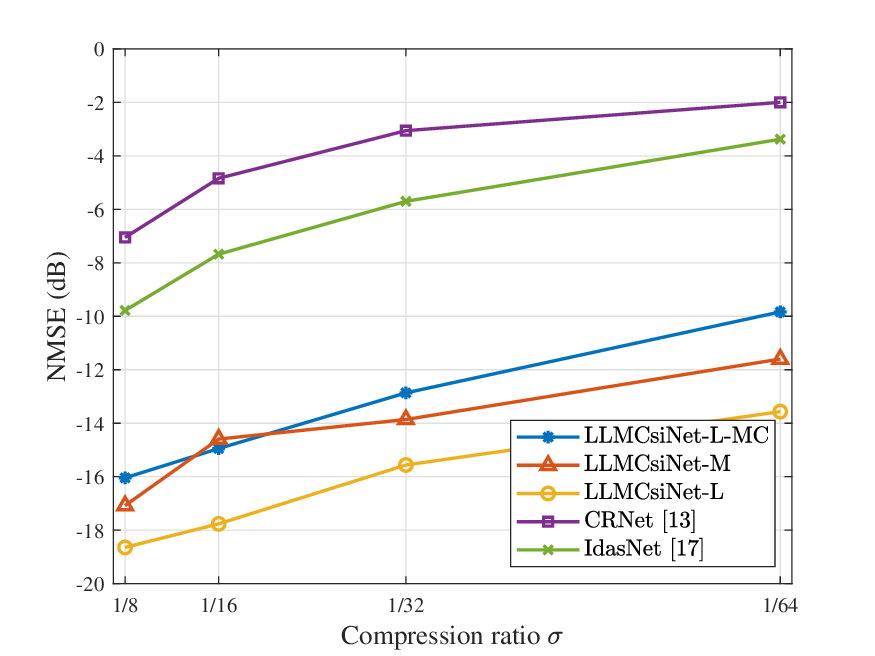}
	\caption{NMSE comparison of multiple compression ratios in the COST2100out channels.}
	\label{multicompress}
\end{figure}

\begin{figure}[t]
	\centering  
	\includegraphics[width=8.5cm]{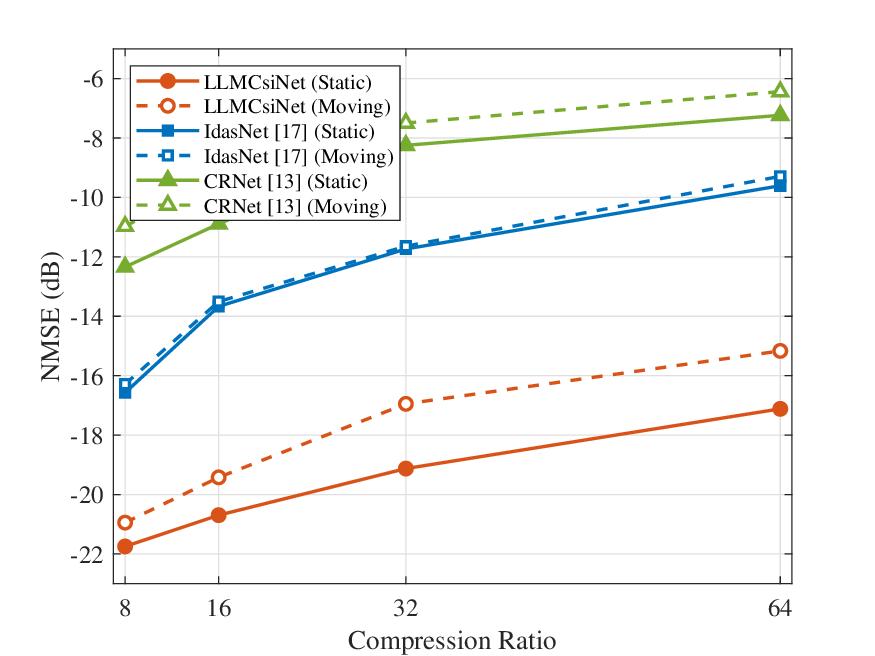}
	\caption{Performance generalization under mobility on the DeepMIMOo1 dataset.}
	\label{moveue}
\end{figure}

\begin{figure}[t]
	\centering 
	\includegraphics[width=8.5cm]{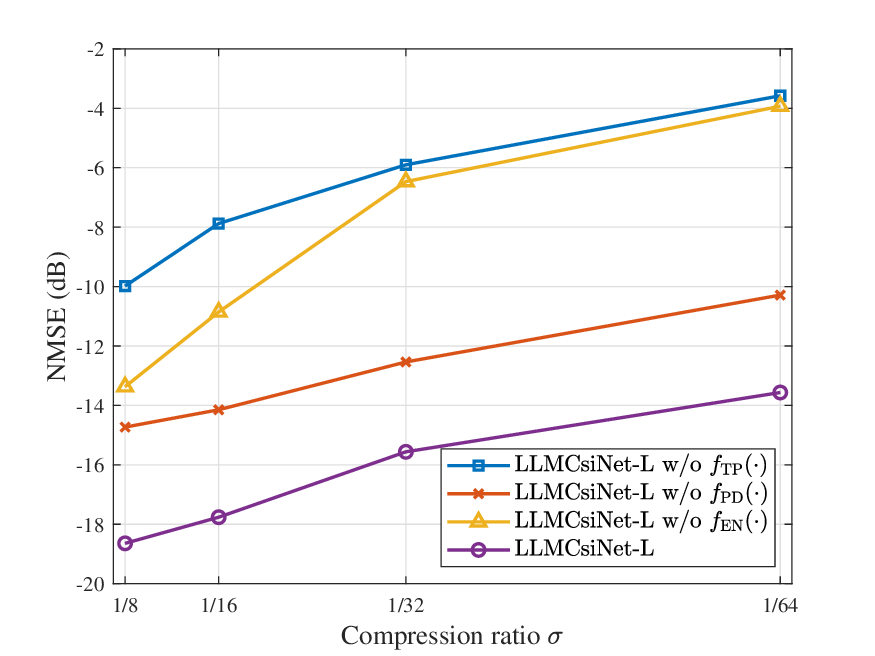}
	\caption{NMSE comparison of the proposed LLMCsiNet-L and three ablation methods in the COST2100out channels.}
	\label{ablation}
\end{figure}
\subsection{Ability to Support Multiple Compression Ratios}

\begin{figure*}[t]
\centering
\subfigure[COST2100in to COST2100out]{
\includegraphics[width=8.5cm]{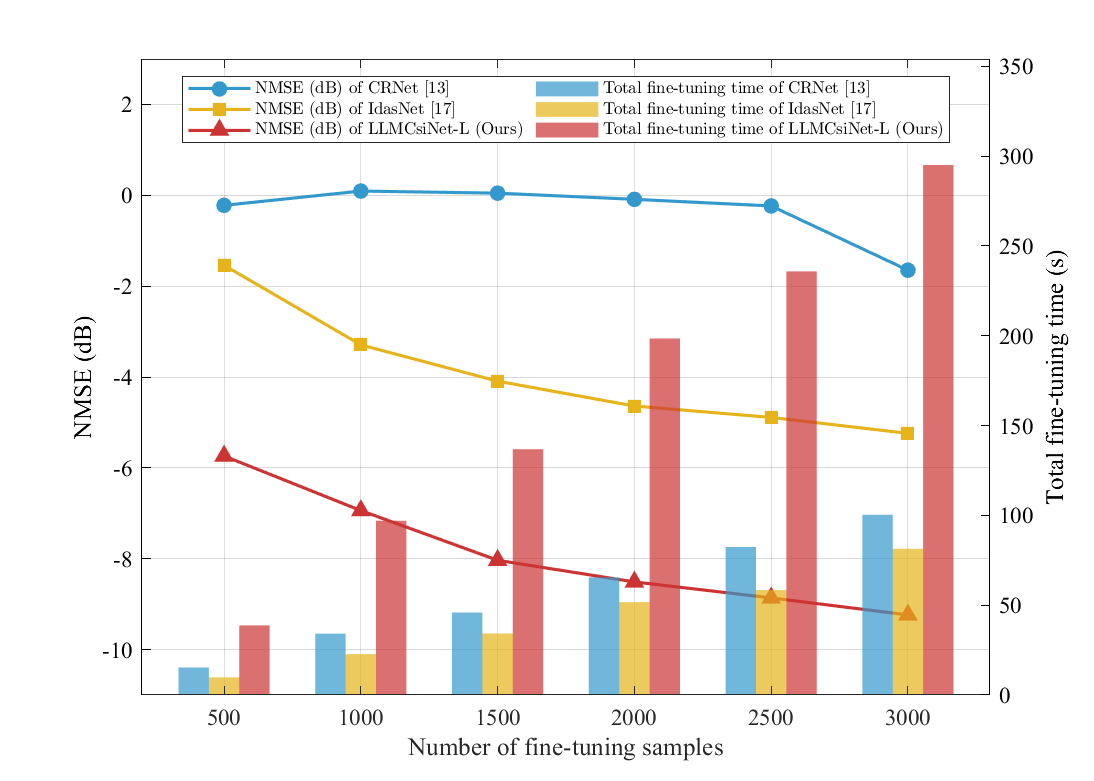}
\label{Fig4-1}
}
\quad
\subfigure[COST2100out to COST2100in]{
\includegraphics[width=8.5cm]{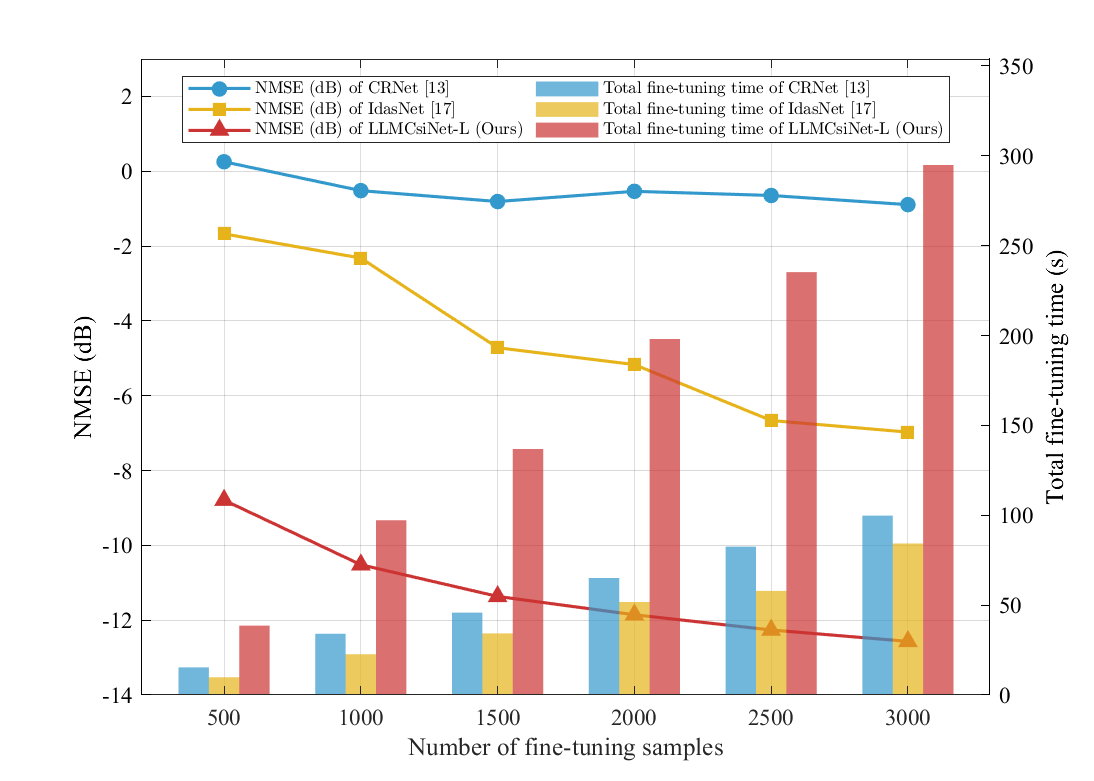}
\label{Fig4-2}
}
\caption{Comparison of NMSE performance and fine-tuning time versus the number of fine-tuning samples.}\label{transfer}
\end{figure*}

Considering that different communication tasks in practical wireless systems often require varying feedback bandwidths, BS must deploy multiple models tailored to different compression ratios in advance. In fact, the proposed LLMCsiNet can be naturally extended to support multi-ratio CSI feedback. This is achieved by allowing the number of high self-information elements selected by $f_{\mathrm{EN}}(\cdot)$ to vary during training rather than being fixed. Specifically, during each training iteration, each CSI sample is randomly compressed to one of the predefined ratios: $1/8$, $1/16$, $1/32$, and $1/64$. We implement this strategy on the LLMCsiNet-L model and denote the resulting multi-ratio capable version as LLMCsiNet-L-MC. We conduct simulation experiments in the COST2100out channels and compare the NMSE performance. The corresponding results are presented in Fig. \ref{multicompress}.

The experimental results clearly demonstrate that a single LLMCsiNet model is capable of efficiently handling CSI reconstruction tasks under multiple compression ratios. Specifically, LLMCsiNet-L-MC achieves consistently superior reconstruction accuracy across the $1/8$, $1/16$, $1/32$, and $1/64$ compression settings, significantly outperforming CRNet and IdasNet. It even matches or surpasses the performance of LLMCsiNet-M. 

\subsection{Generalization Capability for Moving Users}
Considering that user mobility is a fundamental characteristic in practical cellular networks, the CSI feedback mechanism must be robust against channel variations caused by the Doppler effect. To evaluate the generalization capability of LLMCsiNet in dynamic environments, we extend our experiments to include scenarios with moving UEs. Specifically, based on the DeepMIMOo1 dataset, we synthesize user mobility by introducing velocity-dependent Doppler phase shifts to the multipath components, simulating a user speed of 30 km/h. We adopt a mixed training strategy, where the model is trained on a composite dataset consisting of both static and moving channel samples to learn a unified representation. We conduct simulation experiments to compare the NMSE performance against baseline methods, and the corresponding results are presented in Fig. \ref{moveue}.

The experimental results clearly demonstrate that LLMCsiNet possesses strong generalization capabilities to handle CSI reconstruction tasks under user mobility. Specifically, LLMCsiNet exhibits exceptional robustness, showing only marginal performance degradation when transitioning from static to dynamic test sets. It achieves consistently superior reconstruction accuracy across all compression ratios, significantly outperforming CRNet and IdasNet. 

\subsection{Ablation Experiment on Network Modules}

To thoroughly assess the NMSE performance gains introduced by the LLMs within the proposed LLMCsiNet, we conduct ablation experiments using LLMCsiNet-L in the COST2100out channels. Specifically, we remove the modules $f_{\mathrm{EN}}(\cdot)$, $f_{\mathrm{PD}}(\cdot)$, and $f_{\mathrm{TP}}(\cdot)$ to evaluate the individual contribution of each module to the CSI reconstruction performance. The resulting three model variants are denoted as LLMCsiNet-L w/o $f_{\mathrm{EN}}(\cdot)$, LLMCsiNet-L w/o $f_{\mathrm{PD}}(\cdot)$, and LLMCsiNet-L w/o $f_{\mathrm{TP}}(\cdot)$, respectively. In LLMCsiNet-L w/o $f_{\mathrm{EN}}(\cdot)$, the feedback codeword consists of $M$ elements randomly selected from the original CSI, corresponding to a random masking strategy. For LLMCsiNet-L w/o $f_{\mathrm{PD}}(\cdot)$, the codeword produced by $f_{\mathrm{EN}}(\cdot)$ is inserted into a matrix $\mathbf{K}$ constructed from the mean values of the CSI, and the interpolated CSI matrix is directly passed to $f_{\mathrm{TP}}(\cdot)$. In LLMCsiNet-L w/o $f_{\mathrm{TP}}(\cdot)$, the output of $f_{\mathrm{PD}}(\cdot)$ is directly taken as the final reconstructed CSI. Comparison results are shown in Fig. \ref{ablation}.

The simulation results clearly demonstrate that $f_{\mathrm{EN}}(\cdot)$ and $f_{\mathrm{TP}}(\cdot)$ play critical roles in improving the CSI reconstruction accuracy.
When the self-information-based masking module $f_{\mathrm{EN}}(\cdot)$ is replaced with a random masking strategy, the CSI reconstruction accuracy decreases by approximately $5$ dB at high compression ratios and around $10$ dB at low compression ratios. When $f_{\mathrm{TP}}(\cdot)$ is removed, the reconstruction performance of LLMCsiNet-L significantly degrades by approximately $8$ to $10$ dB, comparable to small models. This result highlights the essential contribution of the self-information-based masking strategy and the pretrained LLMs. In addition, removing $f_{\mathrm{PD}}(\cdot)$ also results in a noticeable performance drop, with NMSE degradation ranging from $2$ to $4$ dB under different compression ratios. This indicates that $f_{\mathrm{PD}}(\cdot)$ plays an important auxiliary roles within the overall network architecture. 

\subsection{Comparison of Transfer Learning Abilities}

To rigorously address the critical practical challenge of wireless data scarcity and to benchmark the transferability of the proposed LLMCsiNet, we conduct comprehensive few-shot transfer learning experiments. The experimental setup involves two transfer scenarios between COST2100in and COST2100out at $\sigma=1/8$. The NMSE results, comprehensively depicted in Fig. \ref{transfer}, validate the superior few-shot adaptability of the LLM-based architecture. For instance, in the transfer from COST2100out to COST2100in (Fig. \ref{transfer}(b)), LLMCsiNet-L achieves an NMSE of $-12.57$ dB when fine-tuned with $3,000$ samples, significantly outperforming the best baseline, IdasNet, which only reaches $-6.97$ dB. Furthermore, even with a minimal $500$ samples, LLMCsiNet-L already exceeds the maximum performance attained by IdasNet with the full $3,000$ samples, confirming that the LLMCsiNet effectively achieves high-accuracy reconstruction with minimal data collection effort.

Beyond performance, we analyze the trade-off with computational overhead. While the analysis shows that adapting LLMCsiNet-L takes approximately three times longer than CRNet, this marginal increase in time yields a substantial NMSE improvement ranging from $4$ dB to $7.6$ dB over the baselines. Considering that wireless data collection and labeling are extremely time-consuming and expensive, this modest increase in computation time is highly favorable for rapid and practical deployment.

\subsection{Comparison of the Achievable Rate}
\begin{figure}[t]
	\centering 
	\includegraphics[width=8.5cm]{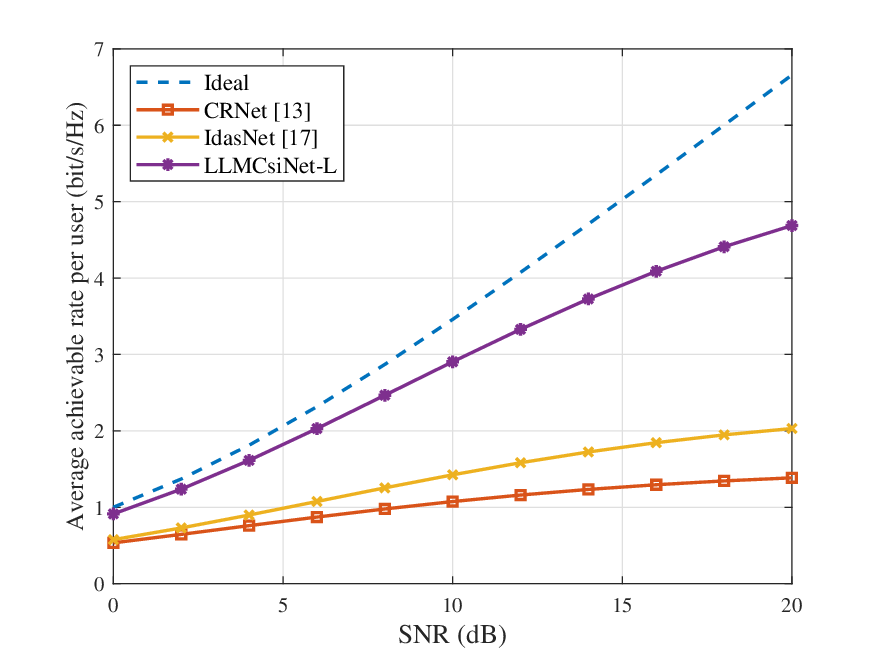}
	\caption{ Comparison of average achievable rate per user performance under the COST2100out channels ($\sigma$ = 1/8).}
	\label{mumimo1}
\end{figure}
In multi-user MIMO scenarios, the achievable rate per user is critically constrained by the accuracy of CSI reconstruction at the BS. Even minor inaccuracies can lead to significant reductions in achievable rates. Under complex channel conditions, traditional small model methods fail to accurately reconstruct CSI, resulting in low achievable rates per user. Our proposed LLMCsiNet, however, provides high-precision CSI reconstruction in complex channel environments, significantly enhancing the achievable rate per user in multi-user MIMO systems. We conduct simulations in the COST2100out channels, with a compression ratio $\sigma$ set to $1/8$ and the number of users set to $4$. Zero-forcing (ZF) precoding is employed for multi-user precoding. 

Simulation results  shown in Fig. \ref{mumimo1} indicate that the proposed LLMCsiNet-L significantly outperforms the small model methods CRNet \cite{deepcsi2} and IdasNet \cite{deepcsi6} under multi-user MIMO scenarios. This demonstrates that achieving high-precision CSI reconstruction through LLMs can overcome the performance bottleneck imposed by CSI reconstruction accuracy in multi-user MIMO systems, providing strong support for the necessity of LLMs in practical communication scenarios.

\section{Conclusion}
In this paper, we propose a novel LLM-driven CSI compression and feedback framework, termed LLMCsiNet. This framework reformulates the CSI compression feedback task into a masked token prediction task that aligns better with LLM functionalities. A self-information-based masking strategy is designed to maximize CSI reconstruction accuracy. Numerical results validate the effectiveness of the proposed LLMCsiNet. Under various channel conditions and compression ratios, LLMCsiNet achieves significantly better NMSE reconstruction accuracy compared to traditional small models, leading to much higher multi-user communication rates. The framework design ensures that additional network complexity is concentrated at the BS, where hardware resources are more abundant, while only a lightweight network is required at the resource-constrained UE. Finally, LLMCsiNet demonstrates excellent generalization performance.

\ifCLASSOPTIONcaptionsoff
  \newpage
\fi

\bibliographystyle{IEEEtran}

\bibliography{IEEEabrv,ref}

\end{document}